\documentclass[longauth]{aa}  
\pdfoutput=1
\usepackage{graphicx}
\usepackage{txfonts}
\usepackage{natbib}
\usepackage{todonotes}

\bibpunct{(}{)}{;}{a}{}{,}

\begin{document} 

\title{Red giants observed by CoRoT and APOGEE:\\
The evolution of the Milky Way's radial metallicity gradient}

\author{
F. Anders\inst{1,2}, C. Chiappini\inst{1,2}, I. Minchev\inst{1}, 
A. Miglio\inst{3}, J. Montalb\'an\inst{4}, B. Mosser\inst{5}, T. S. Rodrigues\inst{2,4,6}, \\ B. X. Santiago\inst{2,7}, F. Baudin\inst{8}, T. C. Beers\inst{9}, L. N. da Costa\inst{2,10}, R. A. Garc\'\i a\inst{11}, 
D. A. Garc\'{i}a-Hern\'{a}ndez\inst{12,13}, \\
J. Holtzman\inst{14}, M. A. G. Maia\inst{2,10}, S. Majewski\inst{15}, S. Mathur\inst{16}, A. Noels-Grotsch\inst{17}, K. Pan\inst{18,15}, D. P. Schneider\inst{19,20}, M. Schultheis\inst{21}, M. Steinmetz\inst{1}, 
M. Valentini\inst{1}, O. Zamora\inst{12, 13}
}

\authorrunning{Anders, Chiappini, et al.}    
\titlerunning{The evolution of the Milky Way's radial metallicity gradient}

\institute{
Leibniz-Institut f\"ur Astrophysik Potsdam (AIP), An der Sternwarte 16, 
14482 Potsdam, Germany
\and{Laborat\'orio Interinstitucional de e-Astronomia, - LIneA, Rua Gal. Jos\'e 
Cristino 77, Rio de Janeiro, RJ - 20921-400, Brazil}
\and{School of Physics and Astronomy, University of Birmingham, Edgbaston, 
Birmingham, B15 2TT, United Kingdom}
\and{Dipartimento di Fisica e Astronomia, Universit\`a di Padova, Vicolo 
dell'Osservatorio 3, I-35122 Padova, Italy}
\and{LESIA, Universit\'{e} Pierre et Marie Curie, Universit\'{e} Denis Diderot, 
Obs. de Paris, 92195 Meudon Cedex, France}
\and{Osservatorio Astronomico di Padova -- INAF, Vicolo dell'Osservatorio 5, 
I-35122 Padova, Italy}
\and{Instituto de F\'\i sica, Universidade Federal do Rio Grande do Sul, Caixa 
Postal 15051, Porto Alegre, RS - 91501-970, Brazil}
\and{Institut d\'{}Astrophysique Spatiale, UMR8617, CNRS, Universit\'{e} Paris XI, B\^{a}timent 121, 91405 Orsay Cedex, France}
\and{Dept. of Physics and JINA-CEE: Joint Institute for Nuclear Astrophysics -- Center for the Evolution of the Elements, Univ. of Notre Dame, Notre Dame, IN 46530 USA} 
\and{Observat\'orio Nacional, Rua Gal. Jos\'e Cristino 77, Rio de Janeiro, RJ 
- 20921-400, Brazil}
\and{Laboratoire AIM, CEA/DSM -- CNRS - Univ. Paris Diderot -- IRFU/SAp, Centre de Saclay, 91191 Gif-sur-Yvette Cedex, France}
\and{Instituto de Astrof\'{\i}sica de Canarias, 38205 La Laguna, Tenerife, Spain}
\and{Universidad de La Laguna, Departamento de Astrof\'{\i}sica, 38206 La Laguna, Tenerife, Spain}
\and{New Mexico State University, Las Cruces, NM 88003, USA}
\and{Department of Astronomy, University of Virginia, PO Box 400325, Charlottesville VA 22904-4325, USA}
\and{Space Science Institute, 4750 Walnut Street Suite  205, Boulder CO 80301, USA}
\and{Institut d'Astrophysique et de Geophysique, All\'ee du 6 ao\^ut, 17 - 
B\^at. B5c, B-4000 Li\`ege 1 (Sart-Tilman), Belgium}
\and{Apache Point Observatory PO Box 59, Sunspot, NM 88349, USA}
\and{Department of Astronomy and Astrophysics, The Pennsylvania State University, University Park, PA 16802}
\and{Institute for Gravitation and the Cosmos, The Pennsylvania State University, University Park, PA 16802}
\and{Observatoire de la Cote d'Azur, Laboratoire Lagrange, CNRS UMR 7923, B.P. 
4229, 06304 Nice Cedex, France}
}

\date{Received 21.07.2016; accepted 20.12.2016}

  \abstract{
Using combined asteroseismic and spectroscopic observations of 418 red-giant stars close to the Galactic disc plane (6 kpc $<R_{\rm Gal}\lesssim13$ kpc, $|Z_{\rm Gal}|<0.3$ kpc), we measure the age dependence of the radial metallicity distribution in the Milky Way's thin disc over cosmic time. 
The slope of the radial iron gradient of the young red-giant population ($-0.058\pm0.008$ [stat.] $\pm0.003$ [syst.] dex/kpc) is consistent with recent Cepheid measurements. For stellar populations with ages of $1-4$ Gyr the gradient is slightly steeper, at a value of $-0.066\pm0.007\pm0.002$ dex/kpc, and then flattens again to reach a value of  $\sim-0.03$ dex/kpc for stars with ages between 6 and 10 Gyr. 
Our results are in good agreement with a state-of-the-art chemo-dynamical Milky-Way model in which the evolution of the abundance gradient and its scatter can be entirely explained by a non-varying negative metallicity gradient in the interstellar medium, together with stellar radial heating and migration. 
We also offer an explanation for why intermediate-age open clusters in the Solar Neighbourhood can be more metal-rich, and why their radial metallicity gradient seems to be much steeper than that of the youngest clusters. Already within 2 Gyr, radial mixing can bring metal-rich clusters from the innermost regions of the disc to Galactocentric radii of 5 to 8 kpc. We suggest that these outward-migrating clusters may be less prone to tidal disruption and therefore steepen the local intermediate-age cluster metallicity gradient. Our scenario also explains why the strong steepening of the local iron gradient with age is not seen in field stars.
In the near future, asteroseismic data from the K2 mission will allow for improved statistics and a better coverage of the inner-disc regions, thereby providing tighter constraints on the evolution of the central parts of the Milky Way.
}


\maketitle

\section{Introduction}

The time evolution of Galactic chemical-abundance distributions is {\it the} missing key constraint to the chemical evolution of our Milky Way \citep[MW; e.g.,][]{Matteucci2003}. 
Several observational questions related to the shape of the present-day abundance distributions as functions of Galactocentric radius, azimuth, height above the disc mid-plane, age, etc., have been tackled in the past (e.g., \citealt{Grenon1972, Luck2003, Davies2009a, Luck2011, Boeche2013, Genovali2013, Genovali2014, Anders2014, Hayden2015, Huang2015}). Especially the radial metallicity gradient, $\partial{\rm [Fe/H]}/\partial R_{\rm Gal}$ -- the dependence of the mean metallicity [Fe/H] of a tracer population on Galactocentric distance $R_{\rm Gal}$ -- has been a subject of debate for a long time.

\begin{figure*}\centering
\includegraphics[width=\textwidth]
{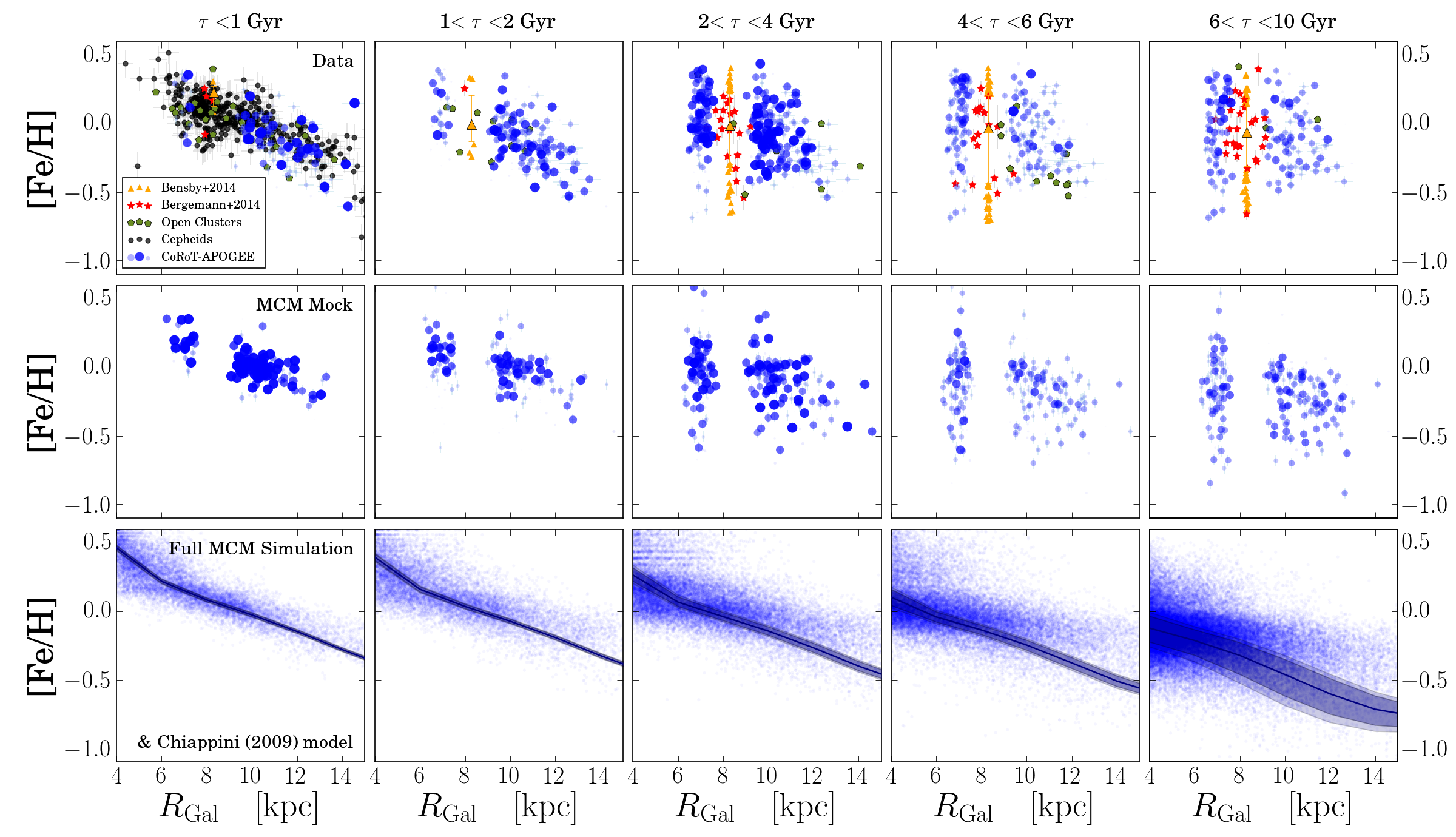}
 \caption{The [Fe/H] vs. $R_{\mathrm{Gal}}$ distribution close to the Galactic plane ($|Z_{\mathrm{Gal}}|<0.3$ kpc) for five bins in age (from left to right, as indicated in each panel). 
{\it Top row:} Data compilation. The CoRoGEE sample is shown in {\it blue}: For each star, we have calculated the fraction of the age PDF which is enclosed in each age bin -- this fraction corresponds to the size of each dot and its transparency value. For comparison, we also show the open cluster compilations of 
\citet{Genovali2014} and \citet{Magrini2015} as {\it green pentagons}, the subgiant sample from \citet[][{\it red symbols}]{Bergemann2014}, and Galactic Cepheids (\citealt{Genovali2014}; {\it black symbols}). The Solar Neighbourhood FGK dwarf sample of \citet{Bensby2014} is plotted as {\it orange symbols}: The big orange triangles and their error bars denote the median metallicities and the 68\% quantiles, while small triangles represent stars which fall outside this range.
{\it Second row:} Mock CoRoGEE sample from the chemodynamical simulation of \citet*[][MCM]{Minchev2013, Minchev2014b}, including typical observational errors in age, distance, and metallicity \citep{Anders2016, Anders2016a}. {\it Third row:} Full MCM simulation without errors, and the underlying chemical-evolution model of \citet{Chiappini2009}.}
 \label{agegradient}
\end{figure*}

Apart from the mathematical representation of the radial metallicity distribution and its dependence on age, also the {\it interpretation} of gradient data remains partly unsettled. Galactic chemical-evolution (GCE) models that reproduce abundance patterns in the Solar Neighbourhood and the present-day abundance gradient can be degenerate in their evolutionary histories (e.g., \citealt{Molla1997, Maciel1999, Portinari2000a, Tosi2000}). One possibility is to start from a pre-enriched gas disc (with a metallicity floor; \citealt{Chiappini1997, Chiappini2001}) which then evolves faster in the central parts, thus steepening of the gradient with time. The other possibility is to start with a primordial-composition disc with some metallicity in the central parts (i.e., a steep gradient in the beginning), and to then gradually form stars also in the outer parts, thus flattening of the gradient with time \citep{Ferrini1994, Allen1998, Hou2000, Portinari2000a}. 
Additionally, radial mixing (through heating or migration, or, most likely, both) flatten the observed abundance gradients of older stars \citep[e.g.][]{Schoenrich2009, Minchev2013, Minchev2014b, Grand2015, Kubryk2015, Kubryk2015a}. Therefore, the gradient of an old population can be flat either because it was already flat when the stars were formed, or because it began steep and was flattened by dynamical processes. 

The answer to the fundamental question of how the MW's abundance distribution evolved with cosmic time is encoded in the kinematics and chemical composition of long-lived stars and can be disentangled most efficiently if high-precision age information is available.
To date, the only results that claim to trace the evolution of abundance gradients are based on planetary nebulae (PNe; e.g., \citealt{Maciel1994, Maciel1999, Stanghellini2006, Maciel2009, Stanghellini2010a}) and star clusters (e.g., \citealt{Janes1979, Twarog1997, Carraro1998, Friel2002, Chen2003, Magrini2009, Yong2012, Frinchaboy2013, Cunha2016}), but the former are based on uncertain age and distance estimates and remain inconclusive\footnote{Also, \citet{Garcia-Hernandez2016} recently showed that the O abundance in PNe (frequently used to study such metallicity gradients) is not an optimal metallicity indicator due to intrinsic O production during the previous asymptotic-giant-branch phase; especially for the more abundant lower-mass-progenitor PNe.}, while the latter allow essentially only for two wide age bins, and are affected by low-number statistics and non-trivial biases due to the rapid disruption of disc clusters.

In addition to the value of the abundance gradient itself, the {\it scatter} of the $R_{\rm Gal}-$[Fe/H] relation can in principle be used to quantify the strength of radial heating and migration over cosmic time. 
In this paper, we use combined asteroseismic and spectroscopic observations of field red-giant stars to examine the evolution of the MW's radial abundance gradient in a homogeneous analysis. Solar-like oscillating red giants are new valuable tracers of GCE, because they are numerous, bright, and cover a larger age range than classical tracers like open clusters (OCs) or Cepheids. The combination of asteroseismology and spectroscopy further allows us to determine ages for these stars with unprecedented precision. 

This paper is structured as follows: In Sec. \ref{obs} we present the data used in this study, and in Sec. \ref{grad} we derive our main result: we present and model the observed [Fe/H] vs. $R_{\mathrm{Gal}}$ distributions in six age bins, and discuss these results in detail in Secs. \ref{dis} and \ref{lit} (the latter focussing on a comparison with the literature). In Sec. \ref{gesexplain}, we revisit the peculiar finding that old open clusters in the Solar Neighbourhood tend to have higher metallicities than their younger counterparts. In Sec. \ref{mggrad}, we analyse the [Mg/Fe] vs. $R_{\mathrm{Gal}}$ distributions, also as a function of age. Our conclusions are summarised in Sec. \ref{concl}.

\section{Observations}\label{obs}

The CoRoT-APOGEE (CoRoGEE) sample \citep{Anders2016} comprises 606 Solar-like oscillating red-giant stars in two fields of the Galactic disc covering a wide range of Galactocentric distance (4.5 kpc $<R_{\mathrm{Gal}}<$ 15 kpc). For these stars, the CoRoT satellite obtained asteroseismic observations, while the Apache Point Observatory Galactic Evolution Experiment (SDSS-III/APOGEE; \citealt{Eisenstein2011, Majewski2015}) delivered high-resolution ($R\sim22,500$), high signal-to-noise ($S/N>90$, median $S/N=240$) $H$-band spectra using the SDSS Telescope at APO \citep{Gunn2006}. The APOGEE Stellar Parameter and Chemical Abundances Pipeline (ASPCAP; \citealt{Holtzman2015, GarciaPerez2016}) was used to derive stellar effective temperatures, metallicities, and chemical abundances; the results are taken from the Sloan Digital Sky Survey's Twelfth data release (DR12; \citealt{Alam2015}). In \citet{Anders2016}, we computed precise masses ($\sim9\%$), radii ($\sim4\%$), ages  ($\sim25\%$), distances  ($\sim2\%$), and extinctions ($\sim 0.08$ mag) for these stars using the Bayesian stellar parameter code PARAM \citep{daSilva2006, Rodrigues2014}, and studied the [$\alpha$/Fe]-[Fe/H] relation as a function of Galactocentric distance, in three wide age bins. Here we slice the data into age bins again, this time examining the dependence of the thin-disc [Fe/H]-$R_{\rm Gal}$ relation on stellar age.

Our final sample comprises 418 stars with $|Z_{\rm Gal}|<0.3$ kpc: 281 are located in the outer-disc field LRa01 $(l,b = 212,-2)$, and 137 in the inner-disc field LRc01 $(l,b = 37,-7)$.\footnote{Our definition of thin disc here is purely geometric, i.e., stars close to the disc mid-plane, as opposed to a definition as the low-[$\alpha$/Fe] sequence in the [Fe/H]-[$\alpha$/Fe] plane (e.g., \citealt{Fuhrmann1998, Lee2011, Anders2014}). While these two definitions agree well at the Solar Radius, they differ especially in the outer disc (see \citealt{Minchev2015} for a discussion on this).}
Because of the location of the CoRoT field LRc01, the cut in $Z_{\rm Gal}$ unfortunately reduces our radial coverage of the inner thin disc, so that we effectively sample Galactocentric distances between 6 and 13 kpc.\footnote{Throughout this paper, we assume $(R_{\rm Gal}, Z_{\rm Gal})_{\odot} = (8.30$ kpc, 0.011 kpc), in line with recent estimates (see, e.g., \citealt{Bland-Hawthorn2016}). All literature data are rescaled to these values.}

As our data exclude the direct Solar Neighbourhood, we complement our analysis by comparing our findings to other recent high-resolution studies focussing on the local Galactic environment: 
For example, \citet{Bensby2014} conducted a spectroscopic Solar Neighbourhood survey of 714 F \& G dwarfs in the Hipparcos volume ($d\lesssim 100$ pc), and derived chemical abundances for 14 chemical 
elements, as well as stellar ages and orbital parameters. We use the 431 stars for which \citet{Bensby2014}, using a kinematical criterion, quote a thin-to-thick-disc probability ratio greater than 1. \citet{Bergemann2014} analysed 144 subgiant stars in an extended Solar Neighbourhood volume (6.8 kpc $<R_{\mathrm{Gal}}<$ 9.5 kpc, $-1.5$ kpc $<Z_{\mathrm{Gal}}< 1.5$ kpc) from the Gaia-ESO survey's first internal data release ({\it i}DR1) to derive accurate ages, Fe, and Mg abundances, and study age-chemistry relations in the MW disc. For this study we select the 51 stars with $|Z_{\rm Gal}|<0.3$ kpc. 

In order to compare our data to the young-population abundance distributions measured with Galactic Cepheid variables, we use the compilation of \citet{Genovali2014}, which comprises spectroscopic iron abundances for several hundred classical Cepheids close to the Galactic mid-plane.

\section{The variation of radial [Fe/H] distributions with age}\label{grad}

Figure \ref{agegradient} shows the age variation of the radial 
metallicity distribution for stars close to the Galactic plane 
($|Z_{\mathrm{Gal}}|<0.3$ kpc), splitting the CoRoGEE stars into five bins in 
age ($0-1$ Gyr, $1-2$ Gyr, $2-4$ Gyr, $4-6$ Gyr, $6-10$ Gyr). The CoRoGEE 
sample is plotted as blue circles, where the size and transparency encode the 
weight, $w_i$, of a star in the age bin considered. For 
instance, a star whose age PDF is fully contained in one age bin appears only 
one time in the diagram, as a large dark blue circle. A star with a broader 
age PDF will appear in multiple panels of the figure, with the symbol size 
and hue in each panel indicating the posterior probability for the star to lie 
in this age bin. 
For comparison, we also plot the radial abundance distribution as measured 
from Galactic Cepheids (black dots; compilation of \citealt{Genovali2014}), 
OCs (green dots; {\it ibid.}) and the 
GES {\it i}DR1 subgiant sample (red dots; \citealt{Bergemann2014}). In order to 
have a better comparison to the Solar Neighbourhood, we further show how the 
FGK dwarf sample of \citet{Bensby2014} is distributed in this diagram. 

\begin{figure}\centering
\includegraphics[width=0.49\textwidth]{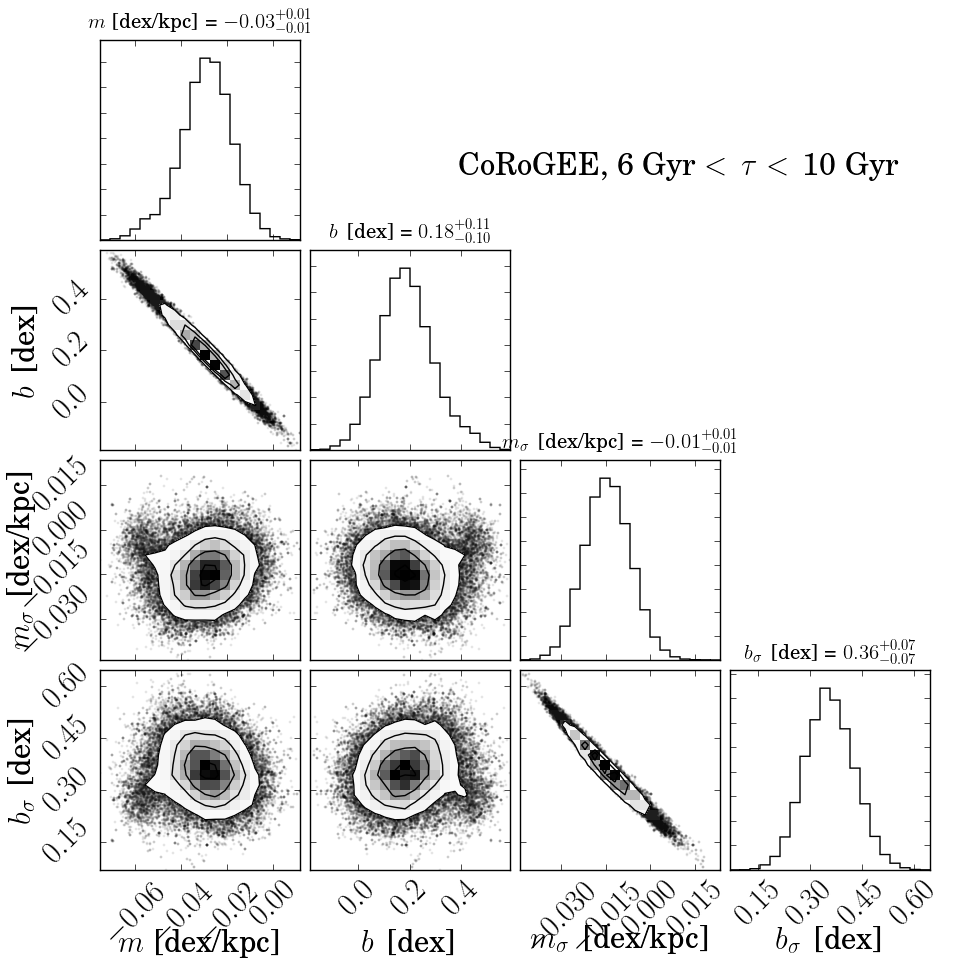}\\
\includegraphics[width=0.49\textwidth]{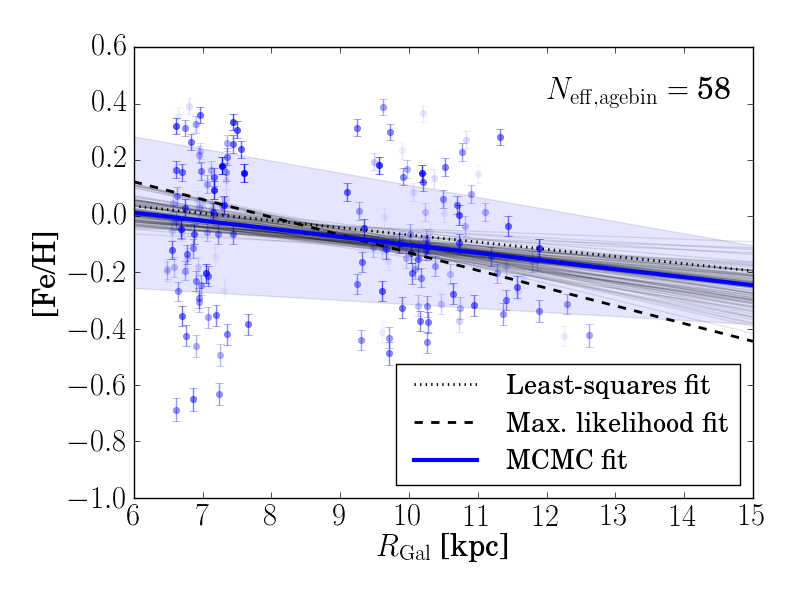}
 \caption{{\it Top panel:} ``Corner plot'' \citep{Foreman-Mackey2016} showing an example of an MCMC fit result for the linear gradient + variable scatter model to CoRoGEE data, in the age bin 6 Gyr $<\tau<$ 10 Gyr. Histograms show the marginal posterior PDFs for the fit parameters $(m, b, m_{\sigma}, b_{\sigma})$; the density-scatter plots show the joint marginal posteriors. {\it Bottom panel:} The resulting fit to the data in the $R_{\rm Gal}$ vs. [Fe/H] diagram. Faint grey lines show 50 MCMC samples; the thick blue line and the shaded band correspond to our best parameter estimates. For comparison, we also show the results of a least-squares linear fit (dotted line) and a maximum-likelihood fit (dashed line).}
 \label{fit_ex}
\end{figure}

The second row of Fig. \ref{agegradient} shows the result of selecting a CoRoGEE-like sample from the N-body chemo-dynamical model of \citet*[][MCM]{Minchev2013, Minchev2014b}. A detailed description of the MCM-CoRoGEE mock sample can be found in \citet[][Sec. 3.2]{Anders2016a}. In short, we select particles from the MCM model that follow the observed $R_{\rm Gal}-Z_{\rm Gal}$ distribution and a red giant-population age bias expected from stellar population-synthesis modelling. We also model the age errors introduced by our observations and statistical inference. 

The third row shows how all the particles from the MCM simulation are distributed in the $R_{\rm Gal}$ vs. [Fe/H] plane, together with the predictions of the pure chemical-evolution thin-disc model of \citet{Chiappini2009}, which was used as an input for the MCM model. The chemical-evolution model was computed in Galactocentric annuli of 2 kpc width, under the assumption of instantaneous mixing within each ring. The bands shown in Fig. \ref{agegradient} reflect the median and the 68\% and 95\% abundance spread within the particular age bin; for this paper we interpolate the model between the $R_{\rm Gal}$ bins. As in \citet{Chiappini2015a}, we scale the abundances of the chemical-evolution model such that the Solar abundances are compatible with the model at the age of the Sun (4.5 Gyr) at the most probable birth position of the Sun (2 kpc closer to the Galactic Centre than today; \citealt{Minchev2013}). This calibration also agrees very well with the abundance scale defined by Galactic Cepheids along the Galactic disc \citep{Genovali2014}.

\subsection{Fitting the ${\rm [Fe/H]}$ vs. $R_{\rm Gal}$ distributions}

From Fig. \ref{agegradient} it is evident that the observed [Fe/H] vs. $R_{\rm Gal}$ distributions should not be fitted with a simple linear model, because the observed scatter is generally larger than the formal uncertainties associated with the measurements. We therefore opt to model the [Fe/H] vs. $R_{\rm Gal}$ distributions for each age bin with a Bayesian ``linear gradient + variable scatter'' model, as follows. 
Any measured metallicity value, [Fe/H]$_i$, is assumed to depend linearly on the Galactocentric radius, $R_{\rm Gal, i}$, convolved with a Gaussian distribution that includes the individual Gaussian measurement uncertainty, $e_{\rm [Fe/H], i}$, and an intrinsic [Fe/H] abundance spread which is allowed to depend linearly on $R_{\rm Gal}$. If we neglect the small uncertainties in $R_{\rm Gal}$ ($\lesssim 2\%$), then the likelihood can be written as:
\begin{footnotesize}
$$ p(\{{\rm [Fe/H]}, e_{\rm [Fe/H]}, R_{\rm Gal}, w\}_i|m,b,m_{\sigma},b_{\sigma}) = \prod_{i=1}^{N} w_i \frac{ \exp(-\frac{({\rm [Fe/H]}_i - {\rm [Fe/H]}_{R,i})^2}{2\sigma_{R,i}})}{\sqrt{2\pi}\sigma_{R,i}^2},$$
\end{footnotesize}
\noindent where 
\begin{align*}
{\rm [Fe/H]}_{R,i} &:= m\cdot R_{\rm Gal, i} + b,\\
\sigma_{R,i} &:= \sqrt{e_{\rm [Fe/H], i}^2 + (m_{\sigma}\cdot R_{\rm Gal, i} + b_{\sigma})^2, }
\end{align*}
\noindent and the weight $w_i$ of each star in a particular age bin is proportional to the integral of its age PDF in this bin:
$$ w_i = \int_{\tau_{\rm min}}^{\tau_{\rm max}} d\tau p(\tau).$$
\noindent
In each age bin, we therefore use an effective number of $N_{\rm eff, agebin} = \sum_i w_i$ stars for the fit. 

We further assume flat priors for each of the fit parameters $m$ (the slope of the radial [Fe/H] gradient), $b$ (its intercept), $m_{\sigma}$ (the slope of the [Fe/H] scatter as a function of radius), and $b_{\sigma}$ (its intercept). The logarithm of the posterior PDF can then be written as:
$$ L_p = -\frac{1}{2}\sum_{i=1}^{N} \left\{ \ln \sigma_{R,i}^2 + \frac{({\rm [Fe/H]}_i - {\rm [Fe/H]}_{R,i})^2}{2\sigma_{R,i}^2} - 2\ln w_i \right\}, $$
\noindent modulo some arbitrary constant. We examine two possibilities: one where all four parameters $(m, b, m_{\sigma},b_{\sigma})$ are allowed to vary, and one where we fix $m_{\sigma}=0$, i.e., an intrinsic [Fe/H] scatter that does not vary with $R_{\rm Gal}$. In the majority of cases the 4-parameter model provides a better fit to the data.

To explore the four-dimensional parameter space and estimate the best-fit parameters and their respective uncertainties, we use the Markov-chain Monte-Carlo (MCMC) code {\it emcee} \citep{Foreman-Mackey2013}. Fig. \ref{fit_ex} shows an example outcome of our fitting algorithm. The grey lines in this figure indicate individual MCMC sample fits, and the blue line and shaded area represent our best-fit (median) results. For comparison, we also show the least-squares and maximum-likelihood results. As in the example case of Fig. \ref{fit_ex}, the three methods generally do not agree within $1\sigma-$uncertainties (especially for the older age bins), which is why we chose the robust Bayesian method (see, e.g., \citealt{Hogg2010, Ivezic2013}).

The results of our attempt to quantify the evolution of the radial [Fe/H] distribution (i.e., fitting for the gradient + scatter) are given in App. \ref{app}, where we provide the tabulated results of our 4-parameter fits to the data gathered in Fig. \ref{agegradient}. 

\begin{figure}\centering
\includegraphics[width=0.49\textwidth]{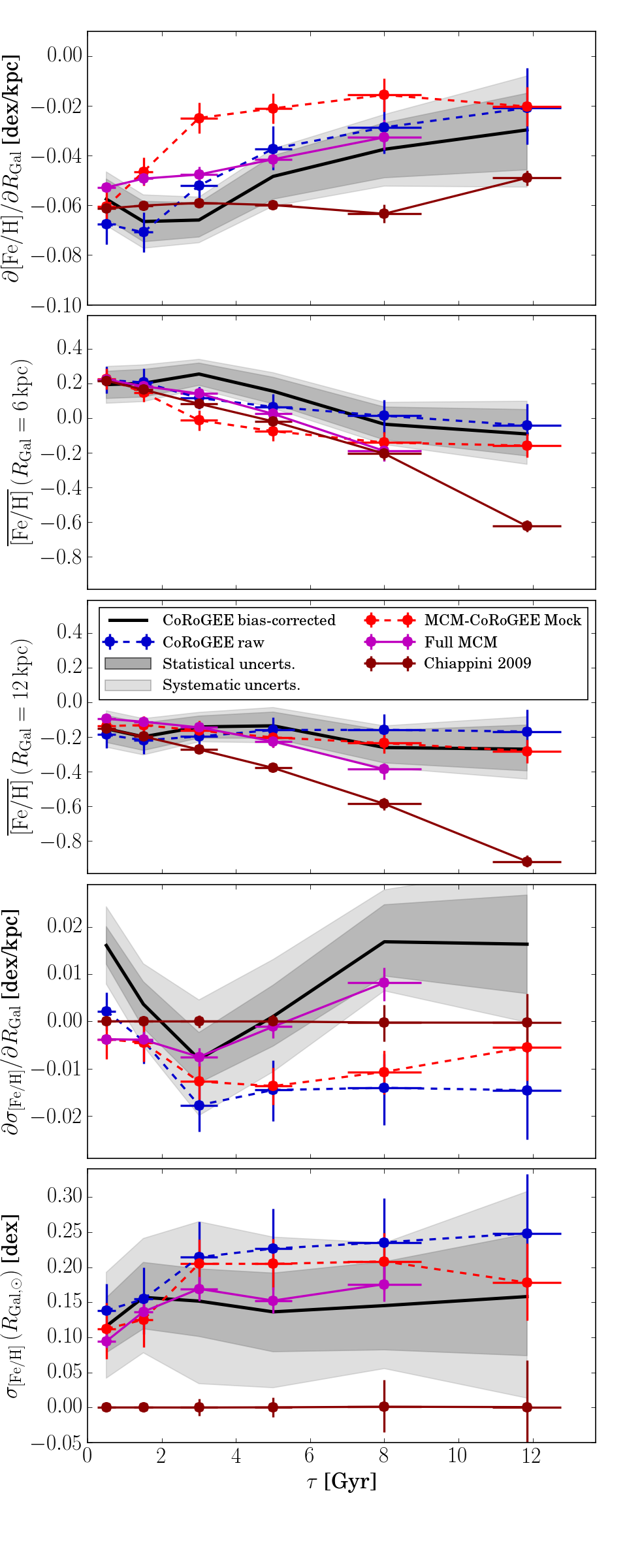}
 \caption{Results of our 4-parameter fits: The evolution of the slope of the radial metallicity gradient between 6 and 13 kpc ({\it top panel}), the mean metallicity at $R_{\rm Gal}=6$ kpc ({\it second panel}), the mean metallicity at $R_{\rm Gal}=12$ kpc ({\it third panel}), the slope of the [Fe/H] scatter with Galactocentric radius ({\it fourth panel}), and the [Fe/H] scatter in the Solar annulus ({\it bottom panel}), for the datasets considered here (error bars). The black line and the shaded bands around it correspond to the bias-corrected CoRoGEE measurements, their statistical uncertainties, and the systematic uncertainties stemming from the bias correction. The fit for the full MCM model did not converge in the last age bin.}
 \label{evolplot}
\end{figure}

Fig. \ref{evolplot} shows the main result of our paper, corresponding to the data compiled in Table \ref{gradienttable_4}: The five panels show the evolution of the [Fe/H] vs. $R_{\rm Gal}$ relation fit parameters with stellar age, for the data and the models used in Fig. \ref{agegradient}. 
The first and the third panel directly display the age-dependence of the fit parameters $m$ and $m_{\sigma}$, respectively, while in the other three panels we show the mean [Fe/H] value at $R_{\rm Gal}=6$ kpc and $R_{\rm Gal}=12$ kpc, respectively, and the [Fe/H] abundance spread in the Solar Neighbourhood -- these latter are linear combinations of the $m$ and $b$, and $m_{\sigma}$ and $b_{\sigma}$, respectively. We discuss the implications of this plot in the next Section.

\section{Discussion}\label{dis}

The CoRoGEE red-giant sample provides an unprecedented coverage of a large range of the Galactic thin disc ($6<R_{\rm Gal}\lesssim13$ kpc, $|Z_{\rm Gal}<0.3$ kpc, 0.5 Gyr $\lesssim\tau\lesssim13$ Gyr) with very precise measurements of [Fe/H], $R_{\rm Gal}$, and ages. 
Although our dataset is not free from biases and selection effects, we can quantify their impact on the derived structural parameters of the MW using mock observations of a chemo-dynamical model that we treat in exactly the same way as the data (Sec. \ref{comp}; see also \citealt{Cheng2012a}). In Sec. \ref{lit} we interpret the results of our analysis presented in Sec. \ref{grad} in the context of past observational literature. We pay special attention to the intriguing old high-metallicity open clusters in the Solar Neighbourhood in Sec. \ref{gesexplain}.
First, however, we remind the reader of the several difficulties that arise when interpreting similar datasets.

\subsection{The importance of fitting earnest(ly)}\label{earnest}

A fit to data can only be as good as the model assumed. The literature (more recently also the astronomical literature) provides good examples of how to find an appropriate model for the dataset considered (e.g., \citealt{Press1992, Feigelson2012, Ivezic2013}). However, in the case of the chemical-abundance gradients, the literature is full of too simplistic approaches. Especially in the case of OCs and PNe, where data are traditionally sparse, biased, and often decomposed further into age bins \citep[e.g.,][]{Janes1979, Twarog1997, Friel2002, Frinchaboy2013, Magrini2015, Jacobson2016}, utmost care has to be taken when fitting a straight line through very few datapoints and drawing conclusions about the evolution of the abundance gradient (see, e.g., \citealt{Carraro1998, Daflon2004, Salaris2004, Stanghellini2010a}). Also, the least-squares method tends to considerably underestimate the uncertainties of the fit parameters when the linear model is inappropriate.

This problem could be mitigated if the number of data points were high enough. Fortunately, this is generally the case for our CoRoGEE sample. In all age bins, the agreement between a least-squares fit and our Bayesian fit for the parameters $m$ and $b$ is at the $1\sigma-$level. Similarly, the effect of including the recent literature data (see Sec. 2) in the fit (i.e., increasing the sample size) is minor ($\lesssim 0.005$ dex/kpc for the gradient), except for the oldest age bin ($\tau>10$ Gyr, $N_{\rm eff}=27$) where the inclusion of literature data increases the sample by a factor of five, and yields a gradient value close to 0 (corresponding to a $+0.02$ dex/kpc with respect to the pure CoRoGEE fit). The fits in this last age bin should therefore be used with caution.

Although our attempt to fit the radial abundance distribution is more elaborate than most methods that can be found in the literature, future data will certainly show if the linear gradient + variable scatter model is sufficient for the part of the Galactic disc considered here. Although there are various indications in the literature \citep[e.g.,][]{Vilchez1996, Afflerbach1997, Twarog1997, Yong2005, Lepine2011, Yong2012} that the radial abundance gradient flattens beyond a break radius $R_{\rm Gal}\sim10-15$ kpc, our combined dataset can be well-fit without a two-fold slope or metallicity step. 
Another important simplification of our model lies in the assumption of Gaussian metallicity distributions at any given Galactocentric distance, which has been shown to be slightly violated in the inner as well as the outer parts of the Galactic disc \citep{Hayden2015}. 

\subsection{Comparison with mock observations of a chemo-dynamical model}\label{comp}

As in \citet{Anders2016}, we opt for the direct approach to compare our observations to CoRoGEE mock observations of a simulated MW-like galaxy which includes all important features of galaxy evolution (merging satellites, disc heating, radial migration, etc.) from high redshift to date.
However, since we want to compare our findings to literature data and other models, in Sec. \ref{corr} we also provide a bias-corrected version of all parameters we measure. The bias corrections were obtained by comparing the mock observations to our fits to the full MCM model. 

\begin{figure}\centering
\includegraphics[width=0.49\textwidth]{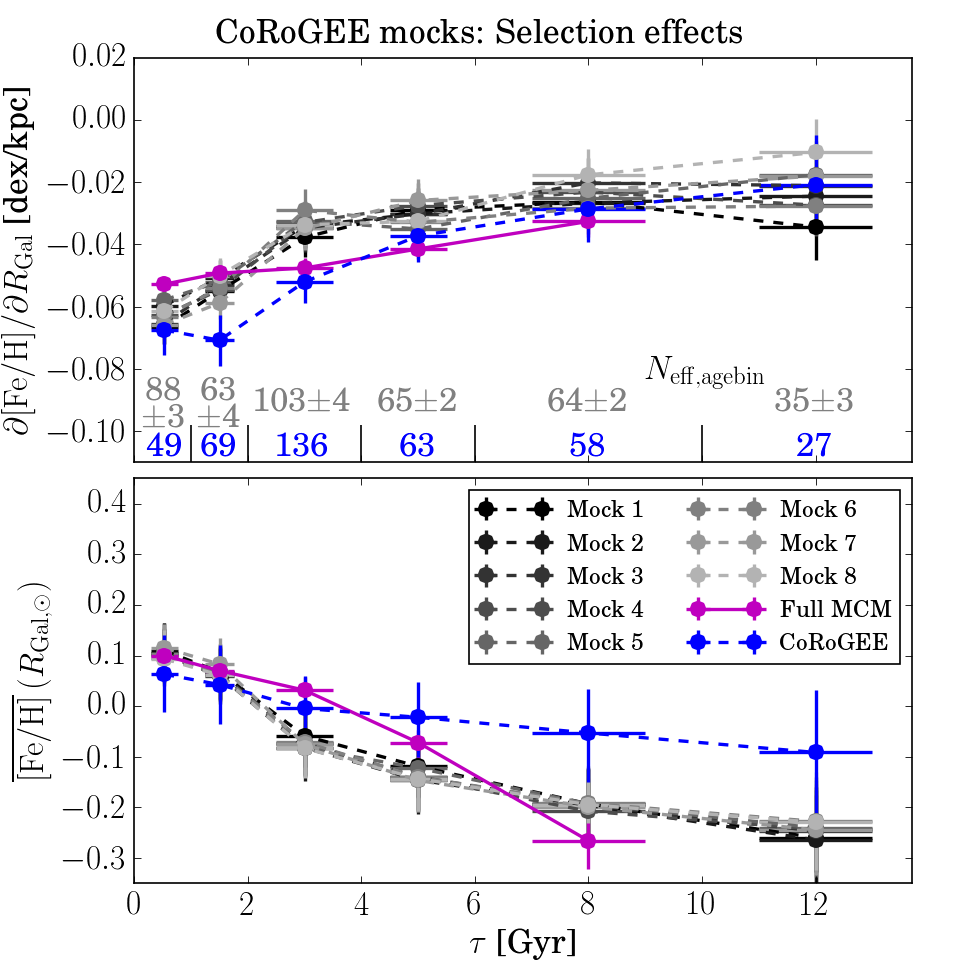}
\caption{Combined effect of mock selection, age errors, and finite sample size on the measurement of the radial [Fe/H] gradients (top panel) and the mean metallicity in the Solar Neighbourhood (bottom), using eight realisations of the MCM-CoRoGEE mock (grey error bars). Also shown are the fit results of the CoRoGEE dataset (blue) and the full MCM model (magenta). The numbers in the top panel indicate the number of stars in each age bin.}
 \label{stoch}
\end{figure}

\subsubsection{From the chemical-evolution model to the MCM model}\label{modmod}

As explained above, the chemo-dynamical model of \citet*[][MCM]{Minchev2013, Minchev2014b} used the pure chemical-evolution thin-disc model of \citet{Chiappini2009} as an input.\footnote{Specifically, the \citet{Chiappini2009} thin-disc model provided the initial birth positions and chemistry for each N-body particle of the cosmological MW-like disc simulation taken from \citet{Martig2012}. The GCE model can thus be seen as the initial Galactic chemistry of the MCM model, which is then mixed by dynamical processes.} In Figs. \ref{agegradient} and \ref{evolplot}, we directly compare our data with these two models. But first, let us recall how the two models {\it intercompare} in those plots.

The four panels of Fig. \ref{evolplot} show the evolution of the main structural parameters of the [Fe/H] vs. $R_{\rm Gal}$ distribution returned by our MCMC fits. Looking at the Galactic models (see also the bottom panels of Fig. \ref{agegradient}), the figure highlights the effects of the realistic Galaxy kinematics included in the MCM model on the [Fe/H] vs. $R_{\rm Gal}$ distributions. With look-back time, radial heating and migration gradually blur out the narrow distributions defined by the thin-disc model, and increase not only the amount of [Fe/H] scatter seen in the Solar Neighbourhood (Fig. \ref{evolplot}, bottom panel), but also wash out the overall gradient (top panel; see also \citealt{Minchev2013}, Fig. 5), and increase the mean metallicity at fixed Galactocentric distance, especially in the outer disc (Fig. \ref{evolplot}, third panel). 

In the semi-analytic GCE model, the width of the [Fe/H] distribution at any fixed radius is entirely due to the finite width of the considered age bin -- i.e., the intrinsic metallicity spread of the model is negligible. In the MCM model, on the other hand, this abundance spread is caused by radial mixing, which already in the $0-1$ Gyr bin widens the [Fe/H] distribution by 0.09 dex in the Solar Neighbourhood (Fig. \ref{stoch}). Already after $2-4$ Gyr, a plateau of $\sigma_{{\rm [Fe/H]}, R_{\rm Gal,\odot}} \simeq 0.15$ dex is reached.

Radial mixing also flattens the chemical gradients predicted by GCE models. In the MCM model, this effect starts to appear for ages $\gtrsim 1$ Gyr: The \citet{Chiappini2009} model predicts a negative radial [Fe/H] gradient of $\frac{\partial {\rm [Fe/H]}}{\partial R_{\rm Gal}} \approx -0.06$ dex/kpc that remains largely constant during the evolution of the MW. While the MCM model predicts an almost unchanged gradient for the very young population ($\tau <1 $ Gyr), the gradients of the older thin-disc populations have flattened to $\sim -0.05$ dex/kpc for $1-4$ Gyr, and to $\sim -0.02$ dex/kpc for $6-10$ Gyr. In the last age bin ($\tau>10$ Gyr), the abundance distribution is so dominated by scatter that it is not well-fit by the linear gradient model any more; the MCMC chains did not converge.  

Radial mixing also affects the mean metallicity in the Solar Neighbourhood and the outer disc -- in the sense that it tends to bring more metal-rich stars from the inner disc into the Solar Neighbourhood than low-metallicity outer-disc stars, because of asymmetric drift and radial migration. This effect starts for ages $\gtrsim 2$ Gyr, and produces a roughly constant mean-metallicity shift on the order of +0.1 dex in the Solar Neighbourhood, and $\sim+0.2$ dex at $R_{\rm Gal}=12$ kpc. 

One effect that is evident from Fig. \ref{agegradient}, but not entirely captured by our linear fitting model, is the emergence of extreme migrators from the inner disc with super-Solar metallicities in the region 5 kpc $<R_{\rm Gal}<8$ kpc. We discuss this result in the context with open-cluster observations in Sec. \ref{gesexplain}.

\subsubsection{From the MCM model to the CoRoGEE mock -- The impact of selection effects}\label{seleff}

In Fig. \ref{stoch}, we assess the impact of selection effects and stochasticity on the measured [Fe/H] gradient: We ran the CoRoGEE mock-selection algorithm \citep{Anders2016a} eight times and fit the linear gradient + variable scatter model to all resulting [Fe/H] vs. $R_{\rm Gal}$ distributions, as before. 
As can be clearly seen in the top panel of Fig. \ref{stoch}, the measured gradient in the CoRoGEE-like MCM mock is varying on the order of $\pm 0.01$ dex/kpc, depending on the realisation of the mock algorithm. The magnitude of these variations are comparable with the uncertainties derived with the MCMC fitting, i.e., our quoted uncertainties are reliable. Since the fit parameters $m$ and $b$ are correlated, the stochasticity effect on the measured mean metallicity is minor.

However, there are also systematic differences between the fit results for the mock realisations and those for the full MCM model, most evidently in the age bins older than 2 Gyr. This is the regime where the combined effects of the red-giant selection bias and systematic age errors begin to matter. 
Specifically, the systematic biases introduced by the selection let the [Fe/H] gradient appear even flatter than in the model, most notably in the age bin 2--4 Gyr, where the effect has an average magnitude of +0.02 dex/kpc. 
A systematic selection effect on the measurement of the mean [Fe/H] near the Sun is also notable in the same age bins: The mock procedure leads to an underestimation of the ``true'' [Fe/H] of the MCM model for the intermediate age bins, and to a slight overestimation in the range 6--10 Gyr. 
In Sec. \ref{corr}, we use these results to correct our CoRoGEE measurements for selection biases, to be able to compare to literature data and other Galactic models.

\subsubsection{Comparing the mock results to the CoRoGEE observations}

The second row of Fig. \ref{agegradient} shows one realisation of the MCM CoRoGEE mock. At first sight, the [Fe/H] vs. $R_{\rm Gal}$ distributions of data and model look remarkably similar, indicating a good overall performance of both the MCM model and the mock algorithm. However, the differences between the relative number of stars in each age bin and field are not within the expected stochastic fluctuations (see numbers in Fig. \ref{stoch}). This mismatch of the age distributions in the mock and the data was already shown in \citet[Fig. 6]{Anders2016a}. We tentatively attribute it to an imperfect modelling of the age bias of the CoRoGEE sample.

Quantitative differences become apparent when looking at the fit results: Fig. \ref{stoch} demonstrates that for $\tau<4$ Gyr, the radial [Fe/H] gradients of all mock realisations are flatter than the observed, and that for ages larger than 2 Gyr the mean [Fe/H] in the Solar Neighbourhood is underestimated by the MCM mocks.

For both model and data, the negative radial metallicity gradient persists in the intermediate-age and old populations, although it gradually becomes shallower and the abundance scatter increases. The increase of this scatter with look-back time is due to a superposition of: a) Growing age uncertainties towards intermediate ages (see  \citealt{Anders2016}); b) Dynamical processes of radial heating and migration \citep[e.g.,][]{Sellwood2002,Schoenrich2009,Minchev2010,Brunetti2011}; and c) Intrinsic time variations of the metallicity distribution (which in the MCM model are negligible; see Sec. \ref{modmod}).
In Fig. \ref{evolplot} we show that the abundance scatter in the MCM model already saturates in the 2--4 Gyr bin, and that almost half of the abundance scatter measured in the subsequent age bins for CoRoGEE may be explained by a combination of abundance and age errors (compare the magenta and red lines).

In the $\tau<1$ Gyr panel of Fig. \ref{agegradient}, the [Fe/H] abundance spread appears to increase somewhat towards very large Galactocentric distances -- an effect that is not seen in the MCM model. The (fewer) data beyond $R_{\rm Gal}\sim 12$ kpc also seem compatible with a flat abundance distribution, as previously argued for in the OC literature \citep[e.g.,][]{Twarog1997, Lepine2011, Yong2012}.
Part of the discrepancies between the data and the MCM model, such as the flattening of the [Fe/H] gradient in the local outer-disc quadrant, may be due to our averaging over the azimuthal angle in Galactocentric cylindrical coordinates (Minchev et al., in prep.).

\subsubsection{Using the mock sample to correct for selection effects}\label{corr}

Under the assumptions that the MCM model is an appropriate model of our Galaxy, and that our mock selection is a good approximation of the true selection function, we can use the differences found between the fits to the MCM mock and the full model (Sec. \ref{seleff}) to correct for the selection biases of the CoRoGEE sample. The first assumption has been extensively tested on a variety of datasets in \citet{Minchev2013, Minchev2014b, Minchev2014c}, and \citet{Piffl2013}. The second assumption was sufficiently validated recently in \citet{Anders2016, Anders2016a}. Of course, the bias correction comes at the price of an additional systematic uncertainty that is driven by the relatively small size of our sample.

\begin{figure}\centering
\includegraphics[width=0.49\textwidth]{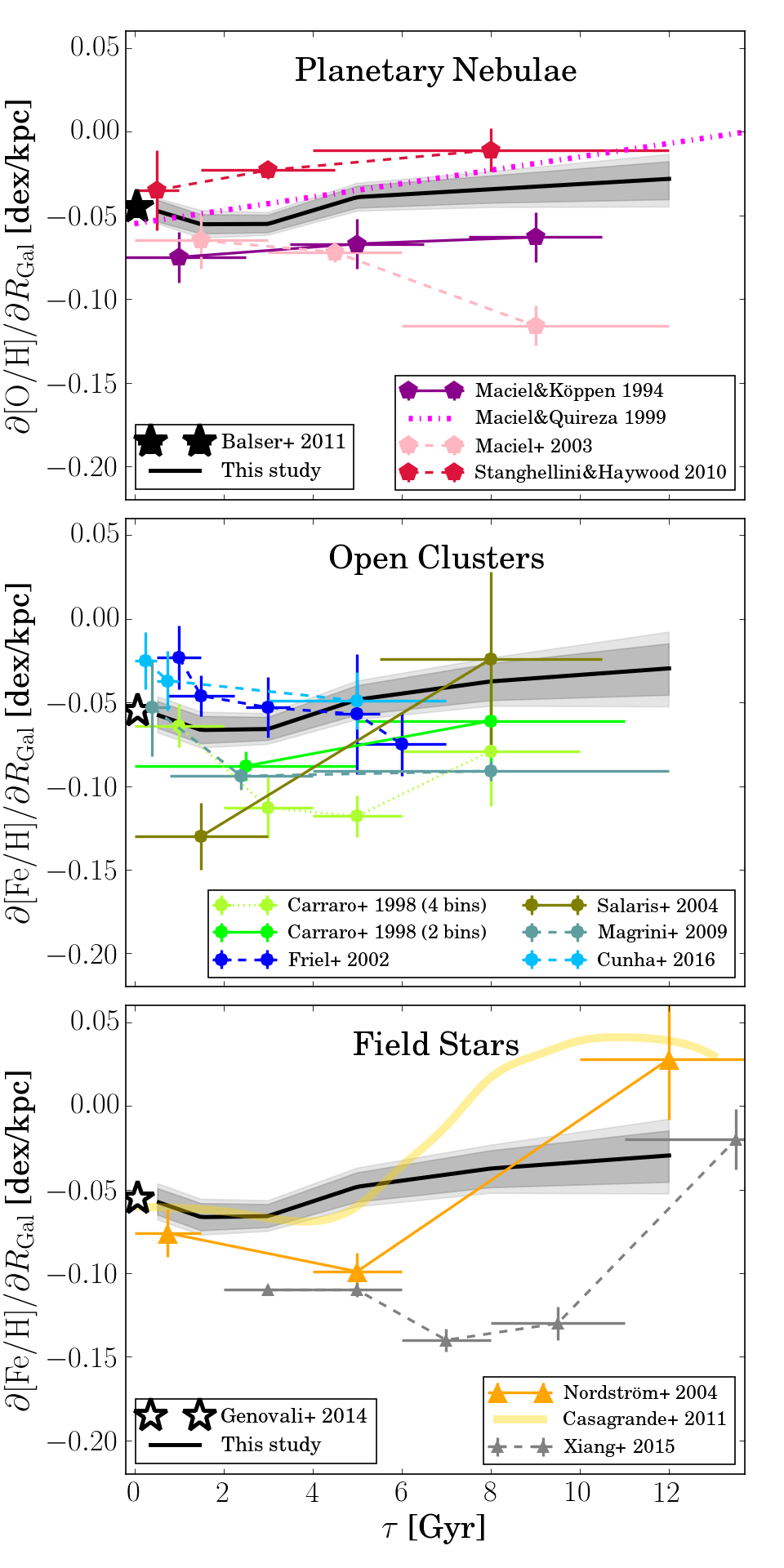}
\caption{An incomplete but representative look into the literature: The dependence of the radial [X/H] gradient slope near the Sun on tracer age $\tau$, as measured by different groups. For the open-cluster ({\it middle panel}) and field-star ({\it bottom}) studies, we plot the results for [Fe/H], for planetary-nebula studies ({\it top}) [O/H]. In all panels, the CoRoGEE results are overplotted as a black line and two grey-shaded bands that correspond to the statistical and systematic uncertainties.}
 \label{history}
\end{figure}

In Figs. \ref{evolplot} and \ref{history}, we also plot the bias-corrected results for the fit parameters measured with CoRoGEE. In each panel the dark-grey bands correspond to the statistical uncertainties of the fit, while the light-grey bands correspond to the systematic uncertainty associated with the bias correction (which we conservatively estimate as the standard deviation of the mock results among different realisations). The additional uncertainty associated with this correction is most important for the fit parameters $m$ and $m_{\sigma}$. 
\footnote{For the oldest age bin, the corrections from the age bin $6-10$ Gyr were used, since we could not perform a reliable fit for the full MCM model in this bin. The corrections for the oldest bin should therefore be used with caution.}

We can now also compare the bias-corrected CoRoGEE results directly to the results obtained from the MCM model. Qualitatively, we reach the same conclusions as before: Overall, the MCM model provides a very good model for our data, and in most of the age bins the fit parameters for data and model are $1\sigma$-compatible (at most $2\sigma$) with each other. As already shown by \citet{Minchev2013, Minchev2014b}, the dynamics of the model give a very good quantitative prescription of the secular processes taking place in the MW. 
The most important differences can probably be fine-tuned in the underlying chemical-evolution model: The observed radial [Fe/H] gradient appears to be slightly steeper than predicted by the model, and the evolution of the mean [Fe/H] from the Solar Neighbourhood inwards is slightly flatter than in the model. Also, at odds with the GCE model, the gradient of the youngest population (in corcodance with Cepheid measurements) seems to be slightly shallower than the gradient of the 1--4 Gyr population. 
However, these discrepancies are marginal compared to other shortcomings of the model, such as the missing bimodality in the [$\alpha$/Fe] diagram, and the systematic uncertainties still involved in stellar age, distance, and abundance estimates.

\subsection{The impact of potential systematic age errors}

Recently, observational as well as theoretical work has raised doubts about the zero-points of the asteroseismic scaling relations that lead to the precise mass and radius measurements for Solar-like oscillating red-giant stars. The absolute scale of seismic masses can only be tested using star clusters and double-lined eclipsing binaries. Recent advances using these techniques for several small datasets \citep{Brogaard2015, Brogaard2016, Miglio2016, Gaulme2016} indicate that asteroseismic scalings tend to overestimate red-giant masses and radii by about +10-15\% and $-5$\%, respectively. A +10\% shift in the seismic masses would result in a $\sim+30\%$ shift of our absolute age scale, and consequently our measurements of the evolution of Galactic structural parameters with stellar age would be affected at this level. However, another potentially important effect that has not previously been taken into account is a metallicity-dependent temperature offset between spectroscopic red-giant observations and stellar models (Tayar et al., in prep.). For PARSEC models, this effect likely {\it decreases} the absolute ages of sub-Solar metallicity CoRoGEE stars by a small amount. Because a systematic analysis of the above effects is still premature, we here restrict ourselves to caution the reader about these caveats to our absolute age scale. For further discussion of systematic age uncertainties, see also Sec. 3 of \citet{Anders2016}.

\section{A comparison with the literature}\label{lit}

As reviewed in the Introduction, the chemical-evolution literature of the past $\gtrsim20$ years has been accompanied by a controversy about both measurement and interpretation of the evolution of the Milky Way's radial abundance gradient. Fig. \ref{history} illustrates the recent history of this controversy: it shows the radial [X/H] gradient as a function of tracer age, as reported by various groups in the literature (compare, e.g., \citealt{Daflon2004, Chiappini2006}).

The datasets as well as the methods used to derive the results included in Fig. \ref{history} are of course very diverse, and deserve a closer look, but the overall situation is clear: There is no consensus about the evolution of the Galactic radial metallicity gradient over cosmic time, not among groups using the same tracers, and sometimes not even among the same groups, or the same datasets. This is essentially due to five reasons: 
\begin{enumerate}
 \item Different radial and vertical ranges of the disc considered;
 \item Different age, distance, and abundance scales among different groups, and between different tracer populations, especially in the case of PNe;
 \item Different selection biases for the various tracers;
 \item Insufficient statistics;
 \item Different fitting methods, handling of outliers, etc. (Sec. \ref{earnest}).
\end{enumerate}

The literature results included in Fig. \ref{history} were derived using four different tracers (PNe, OCs, Cepheids, and low-mass field stars) that are discussed separately below. For the radial [Fe/H] gradient traced by the very young population (white star in Fig. \ref{history}) we use the recent result of \citet{Genovali2014}, which is based on a sample of 450 Galactic Cepheids located at Galactocentric distances between 5 and 12 kpc, using data from \citet{Lemasle2007, Lemasle2008, Romaniello2008, Luck2011, Luck2011a, Genovali2013}. The authors find a gradient slope of $-0.05$ to $-0.06$ dex/kpc, slightly depending on the adopted cuts in $R_{\rm Gal}, |Z_{\rm Gal}|$. We therefore adopt a value $-0.055\pm0.05$ dex/kpc, which coincides with their value for $|Z_{\rm Gal}|<0.3$ kpc. For the [O/H] gradient, other young tracers such as OB stars or HII regions have to be used \citep{Deharveng2000, Daflon2004, Rood2007, Balser2011}; those studies report slightly flatter slopes ($\sim-0.04\pm0.01$ dex/kpc). In Fig. \ref{history}, we include as a representative datum the value $-0.045\pm0.005$ dex/kpc, obtained by \citet{Balser2011} from high-quality radio observations of 133 HII regions covering a large azimuthal range of the Galactic disc.

The PNe-based results for the evolution of the radial [O/H] gradient are shown in the top panel of Fig. \ref{history}. We have used the same technique as before to also fit the radial [O/H] distribution (results can be found in table \ref{gradienttable_oh_4}), and compare the resulting gradient slope to three representative values from the literature. The studies by \citet[][magenta pentagons]{Maciel1994a} and \citet[][red pentagons]{Stanghellini2010a} divided their objects into the three classical categories of disc PNe, type I, II, and III \citep{Peimbert1978, Faundez-Abans1987}, corresponding to different masses and hence ages of the PNe's central stars. In their paper, \citet{Maciel1994a} find hints that the radial [O/H] gradient flattens slightly with look-back time, starting from a steep value of $-0.075\pm0.015$ dex/kpc for massive type-I PNe to $-0.063\pm0.015$ for the low-mass type-III objects. \citet{Maciel1999} reach a similar conclusion and, extrapolating the trend to the early phases of the Galactic disc, calculate a rough estimate of the temporal flattening of the [O/H] gradient with age: $\frac{\partial}{\partial \tau} \frac{\partial{\rm [O/H]}}{\partial R_{\rm Gal}} \sim -0.004\, {\rm dex\, kpc^{-1}\, Gyr^{-1} }$.
This estimate is in overall agreement with the trend we derive from the CoRoT-APOGEE data. \citet{Stanghellini2010a} also find that the radial [O/H] gradient flattens with look-back time, although their derived slopes are generally flatter.

However, the early findings of Maciel and collaborators are at odds with their later results (e.g., \citealt{Maciel2003, Maciel2009, Maciel2012}) that report an opposite trend for the evolution of the radial [O/H] gradient. These later studies preferred different methods to assign PN ages over the traditional Peimbert classification scheme. For example, \citet[][light-pink pentagons in Fig. \ref{history}]{Maciel2003} derived ages using statistical relationships between [O/H] and [Fe/H], as well as an age-metallicity-$R_{\rm Gal}$ relationship, which makes their subsequent measurement of the radial metallicity gradient using these ages a rather circular exercise. 

Because the PN studies still yield inconclusive results, we caution against arbitrary use of PN-derived abundance gradients (see also \citealt{Stasinska2004, Stanghellini2006, Stanghellini2010a, Garcia-Hernandez2016} for additional remarks regarding the derivation of PN abundances and distances).

In Fig. \ref{history}, we also show several results derived from spectroscopic OC observations \citep{Carraro1998, Friel2002, Salaris2004, Magrini2009, Cunha2016}. This is by no means an exhaustive compilation, but the diversity of the conclusions reached is representative. As pointed out in Sec. \ref{earnest}, many of these results rely on very few datapoints, so that their linear fits are subject to significant influence by single outliers (e.g., Berkeley 29 and NGC 6791; see also Sec. \ref{gesexplain}), and sometimes the OC samples are dominated by outer-disc objects. 

The 37 clusters studied by \citet[][lime-green points in Fig. \ref{history}]{Carraro1998} span a wide range of Galactocentric distances ($7-16$ kpc), but also a considerable range of heights above the Galactic-disc plane ($0-2.1$ kpc). This makes their results difficult to compare, as the clusters located at high $|Z_{\rm Gal}|$ also tend to be at larger $R_{\rm Gal}$, i.e. it is almost impossible to disentangle radial and vertical trends, and their quoted radial [Fe/H] gradients are therefore very likely to be significantly overestimated for the older age bins. Also, the authors demonstrate that the age binning impacts their conclusion about the evolution of the abundance gradient along the disc (compare dotted and solid lime-green curves: 4 bins vs. 2 bins), because of the lack of OCs older than $>2$ Gyr. 

\begin{figure*}\centering
\includegraphics[width=0.49\textwidth]{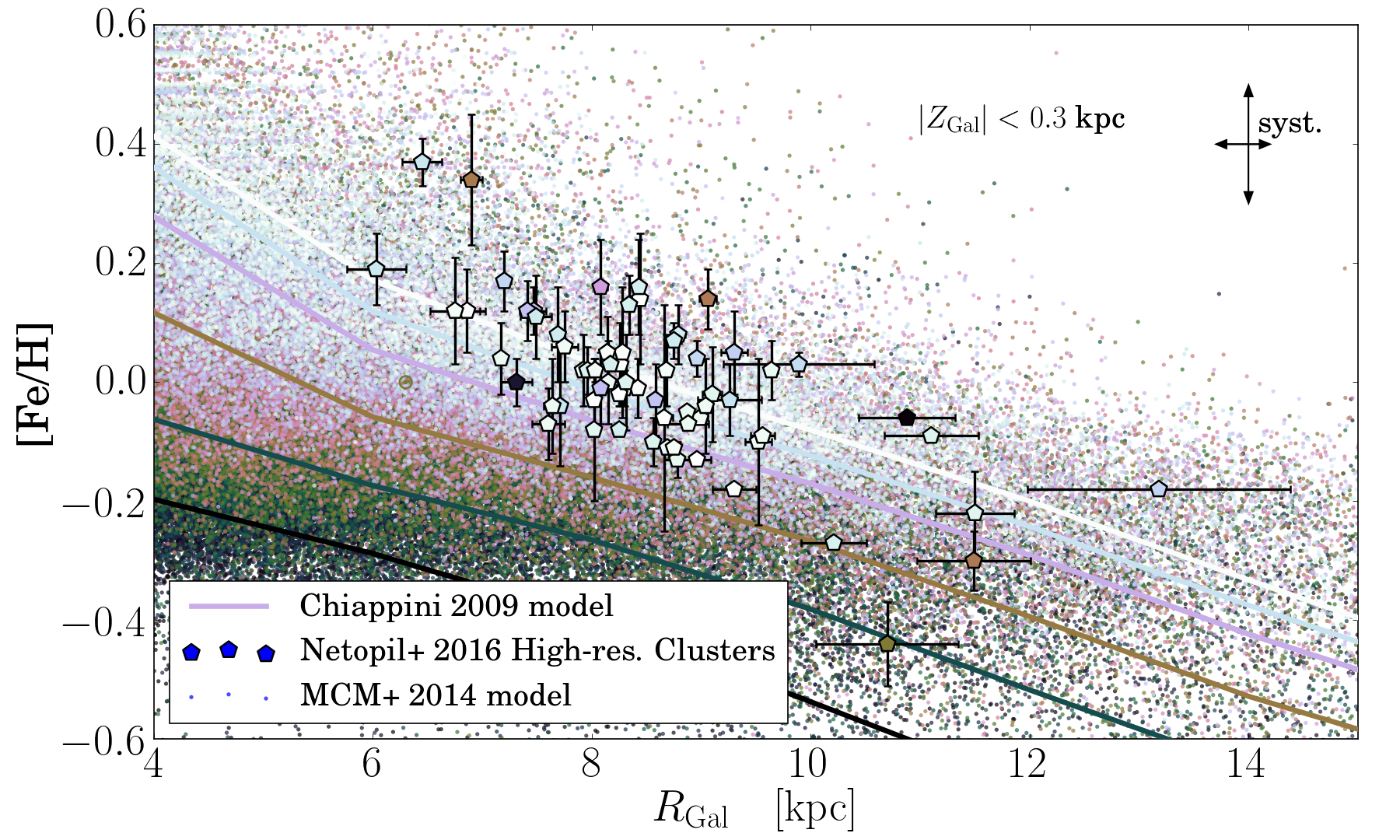}
\includegraphics[width=0.49\textwidth]{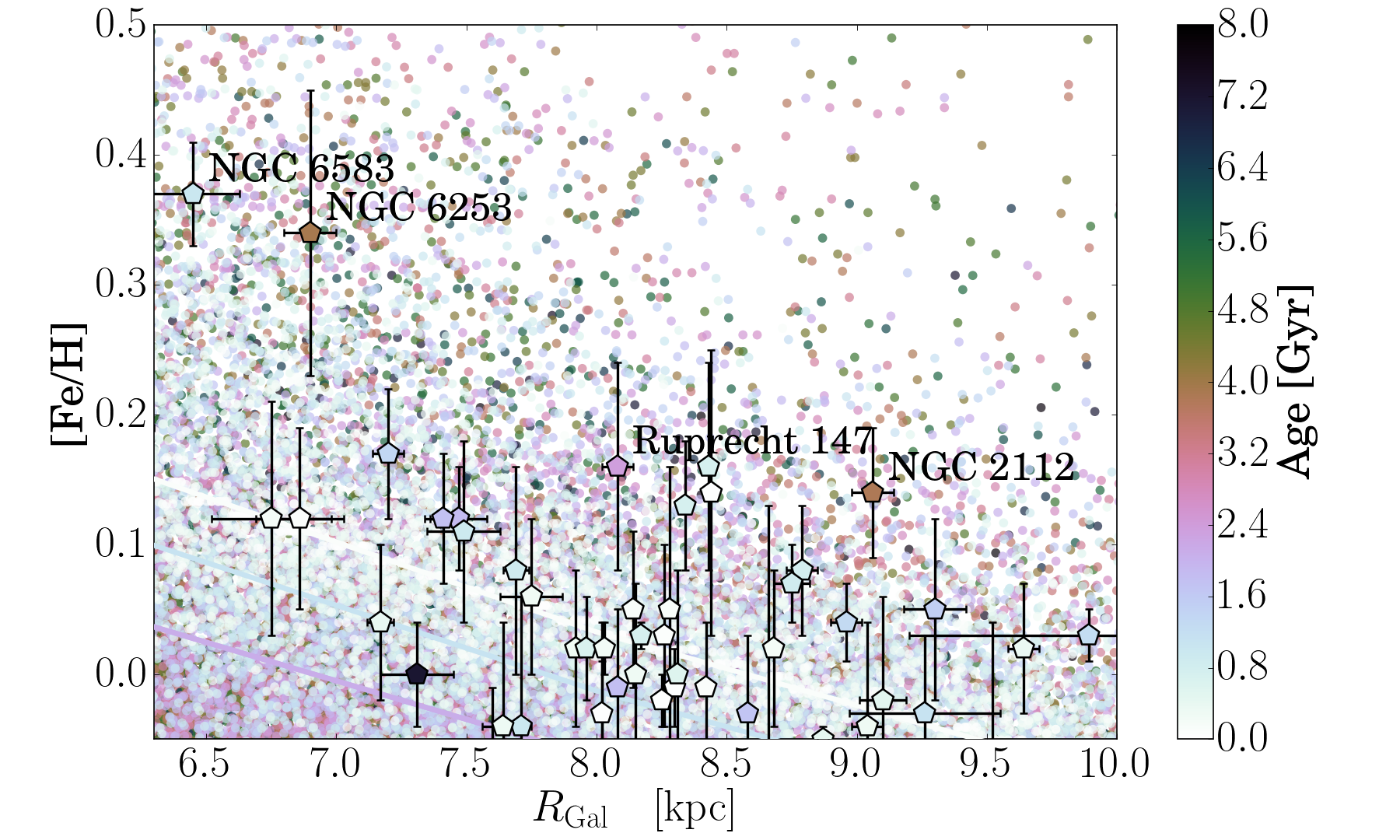}
 \caption{The [Fe/H] vs. $R_{\rm Gal}$ distribution, colour-coded by age, for the high-resolution OC compilation of \citet[][{\it pentagons}]{Netopil2016}, the MCM model close to the Galactic plane ($|Z_{\rm Gal}|<0.3$ kpc; {\it small dots}), and the \citet{Chiappini2009} thin disc model at six time snapshots ({\it thick lines}). The models have now been scaled to the abundance scale of young OCs in the Netopil et al. sample, which is $\lesssim0.1$ dex lower than the Cepheid abundance scale of \citet{Genovali2014} used before. The arrows in the left panel symbolise the approximate uncertainties in the absolute metallicity scale and the distance to the Galactic Centre. The {\it right panel} magnifies the interesting part of the left panel, such that it is clear that the some older OCs have higher metallicities than both the younger ones and the local ISM. With the MCM model, these can be explained as surviving migrators from the inner disc.}
 \label{clusterplot}
\end{figure*}

This situation has improved slightly in the past few years; e.g., in the works of \citet{Friel2002} and \citet{Chen2003} the number of chemically-studied OCs increased to over 100, and \citet{Friel2002} specifically focussed on old OCs to trace the evolution of the Galactic radial metallicity gradient. Overall, the OC works of \citet{Friel2002, Chen2003, Magrini2009}, and also the later studies of \citet{Frinchaboy2013} and \citet{Cunha2016}, using APOGEE data, all reach the same conclusions: that the radial metallicity gradient has been steeper in the past. However, this paradigm has been challenged by \citet{Salaris2004} who, using the old OC sample of \citet{Friel1995}, reach the opposite conclusion, namely that the radial [Fe/H] gradient of the old OC population is shallower than the gradient of the young population. In summary, and similar to the case of the PN studies (but due to different reasons), OC studies still fail to conclusively answer the question of the evolution of the abundance gradient along the MW disc. 

Some works based on field stars attempted to measure the age-dependence of the Galactic radial metallicity gradient, all of them derived from low-resolution spectroscopic stellar surveys. The Geneva-Copenhagen survey of the Solar Neighbourhood (GCS; \citealt{Nordstroem2004}, orange triangles in Fig. \ref{history}) derived kinematics, metallicities, and ages for 14,000 FGK dwarfs for which precise astrometric distances were delivered by the {\it Hipparcos} satellite \citep{Perryman1997, vanLeeuwen2007}. Because their sample does not extend beyond the immediate Solar Neighbourhood, the authors used the precise kinematic information to correct for the eccentricity of stellar orbits, and report metallicity distributions as a function of orbital guiding radii $R_g$ (instead of $R_{\rm Gal}$), in three bins of age. While their result for the young population ($\tau<2$ Gyr) is consistent with our findings, their intermediate-age population (4 Gyr $<\tau<$ 6 Gyr) exhibits a steeper negative gradient on the order of $-0.1$ dex/kpc (possibly related to the fact that they use $R_g$ as a baseline), and the old population is consistent with no radial gradient.

The age estimates derived in \citet{Nordstroem2004} were later improved \citep{Holmberg2007, Holmberg2009, Casagrande2011}, and their results for the age dependence of the metallicity gradient were also revised by \citet{Casagrande2011}. In the last panel of Fig. \ref{history}, we plot the smooth result that these authors obtain for the evolution of the radial metallicity gradient with respect to the mean orbital radius, computing the gradient in a Gaussian age window of 1.5 Gyr width, and using a kinematic cut to sort out halo stars. Because their sample (even in a given age window) is very large, the formal uncertainties on the gradient measurement are almost negligible. 
Again, the GCS data seem to indicate an initial steepening, then a rapid flattening with increasing age, although \citet{Casagrande2011} caution that this picture depends severely on selection effects. They invoke radial migration to explain the observed softening of the gradient at intermediate ages. 

The main advantage of the Solar-Neighbourhood studies of \citet{Nordstroem2004} and \citet{Casagrande2011} lies in their use of very precise kinematic parameters in combination with unprecedented statistics, an asset that will soon be provided for a much larger volume by the {\it Gaia} mission \citep{Perryman2001}. On the other hand, the GCS metallicities and -- above all -- ages are likely to be affected by significant systematic shifts, and contamination by thick-disc stars. 

The other stellar survey that recently led to an estimation of the age-dependence of the radial metallicity gradient is the LAMOST Spectroscopic Survey of the Galactic Anti-Center (LSS-GAC; \citealt{Liu2014}). \citet{Xiang2015} have used $\sim 300,000$ main-sequence turn-off stars selected from that survey to measure the radial and vertical stellar metallicity gradients as a function of stellar age, which they determine via isochrone fitting. With their impressive statistics, these authors can measure very precise relative stellar-population trends. After correcting for selection effects, they find that the radial [Fe/H] gradient close to the Galactic plane steepens with age until $\tau\sim 7-8$ Gyr before flattening again, and interpret these time spans as corresponding to two distinct phases of the assembly of the MW disc.

However, the analysis of \citet{Xiang2015} is based on low-resolution, low-signal-to-noise data, leading to much larger individual uncertainties in their ages, distances, and metallicities compared to our data. Although this is partly mitigated by the large number of stars, considerable systematic uncertainties and poor absolute calibrations are expected for their age and distance estimates, so that we do not expect the absolute values of their reported gradients to perfectly match ours, nor other literature values.

Recently, several authors \citep{Anders2014, Hayden2014, Bovy2014b} have used data from the APOGEE survey to measure the radial disc metallicity gradient. Our bias-corrected values for the gradient slope are slightly shallower than theirs, but compatible with them within our quoted uncertainties. The differences in the absolute values of the metallicity gradient are most probably due to different selection biases in previous works. For example, \citet{Bovy2014b} measured a slope of $-0.09$ dex/kpc based on 971 RC stars very close to the plane ($Z_{\rm Gal} < 50$ pc), while our sample extends up to distances of 300 pc from the Galactic mid-plane. The APOGEE RC sample is also much more limited in age coverage ($\sim 0.8-4$ Gyr with a peak close to 1 Gyr, see e.g. \citealt{Girardi2016}) than our sample. In contrast to the results of the above authors, and because our sample is considerably smaller, we have corrected our results for selection biases and included a corresponding systematic uncertainty.

\section{Intermediate-age high-metallicity open clusters in the Solar Neighbourhood}\label{gesexplain}

Radial mixing (by migration and heating) can explain the presence of super-metal-rich stars in the Solar Neighbourhood \citep[e.g.,][]{Grenon1972, Grenon1999, Chiappini2009, Minchev2013, Kordopatis2015, Anders2016}. 
Here we show that radial migration can also explain the existence of intermediate-age super-solar metallicity objects in the solar-vicinity thin disc ($R_{\rm Gal}=7-9$ kpc). In the MCM model, those objects originate from the inner disc ($R_{\rm Gal}=4-6$ kpc) and already start to appear in the solar vicinity after 2--4 Gyr after their birth (see Fig. \ref{agegradient}, bottom row). In this Section we demonstrate this for the case of intermediate-age open clusters in the solar vicinity.

The left panel of Fig. \ref{clusterplot} shows the [Fe/H] vs. $R_{\rm Gal}$ distribution of the homogenised OC compilation recently published by \citet{Netopil2016}, colour-coded by cluster age. The right panel zooms closer into the Solar Neighbourhood. For all of the clusters included in the plot, [Fe/H] was derived from high-resolution spectroscopy, in most cases by different groups (see \citealt{Netopil2016} and references therein for details). 
In Fig. \ref{clusterplot} we also show the MCM model for $|Z_{\rm Gal}|<0.3$ kpc and the thin-disc model of \citet{Chiappini2009}, for snapshots at $\tau= 0, 1, 2, 4, 6,$ and 8 Gyr. Because we found the Fe abundance scales of OCs and Cepheids to be slightly offset with respect to each other ($\lesssim 0.1$ dex), for this plot the models' [Fe/H] values were rescaled to match the abundance scale defined by the youngest OCs of \citet{Netopil2016} in the Solar Neighbourhood. 

It can be clearly seen from Fig. \ref{clusterplot} that, while the chemical-evolution model alone cannot explain the location of all of the clusters in the [Fe/H] vs. $R_{\rm Gal}$ plane, the MCM model can: For each cluster there is an N-body particle in the model with very similar properties $\{R_{\rm Gal}, {\rm [Fe/H]}, \tau\}$. In particular, the MCM model predicts that slightly older clusters can be found at larger metallicities than the youngest ones, in the $5-8$ kpc region of the disc (see also \citealt{Netopil2016}). Contrary to the interpretation of \citet{Jacobson2016}, this effect is entirely due to radial mixing, and does not mean that the local ISM had a higher metallicity in the past, nor that the radial [Fe/H] gradient was significantly steeper in the past. 

We remind the reader that the fact that for each OC we find a model particle with similar properties does not mean that the model fits perfectly: The selection function (as a function of age and position) of any OC catalogue would be required for any meaningful comparison of the {\it distributions} of model and data points in Fig. \ref{clusterplot}. This, in turn, requires knowledge about the disruption timescales of clusters as a function of their masses and kinematics, survival rates, initial mass functions, etc. that still remain uncertain. The absence of lower-metallicity intermediate-age OCs in the inner disc (Fig. \ref{clusterplot}, left panel) especially suggests that non- or inward-migrating OCs may be more prone to disruption. This would lead to a higher rate of high-metallicity OCs in the $R_{\rm Gal}=6-8$ kpc regime, and consequently a significantly steeper (up to a factor of 2) local gradient of the intermediate-age OC population, as measured in the OC literature (e.g., \citealt{Carraro1998, Friel2002, Jacobson2016}). The prediction of the MCM model is that this does not happen for the general field-star population of the same age, which is confirmed by the CoRoGEE data.

\begin{figure*}\centering
\includegraphics[width=\textwidth]
{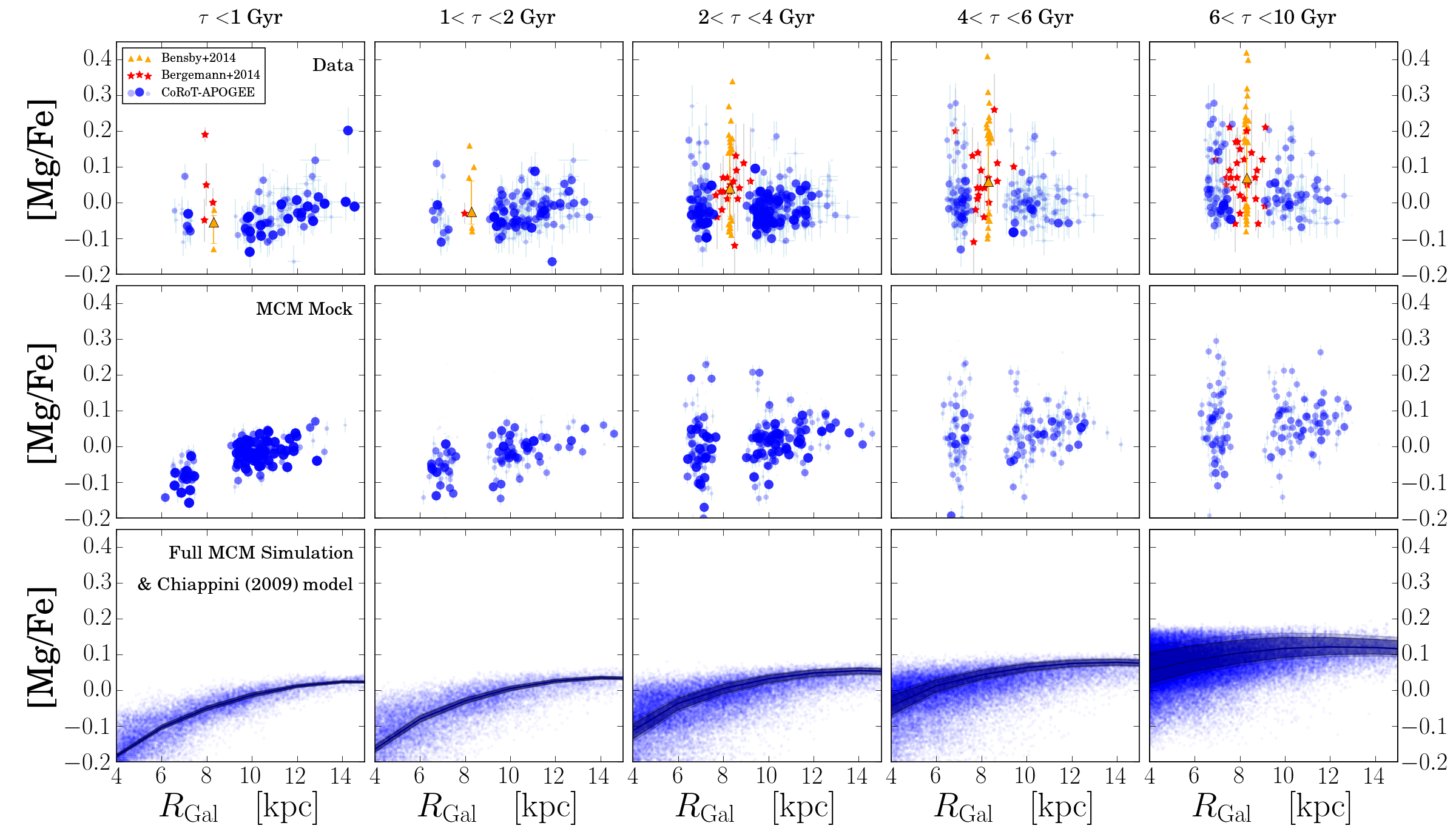}
 \caption{The [Mg/Fe] vs. $R_{\mathrm{Gal}}$ distributions close to the Galactic plane ($|Z_{\mathrm{Gal}}|<0.3$ kpc), in the same style as Fig. \ref{agegradient}.}
 \label{mgfegradient}
\end{figure*}

\begin{figure}\centering
\includegraphics[width=0.49\textwidth]{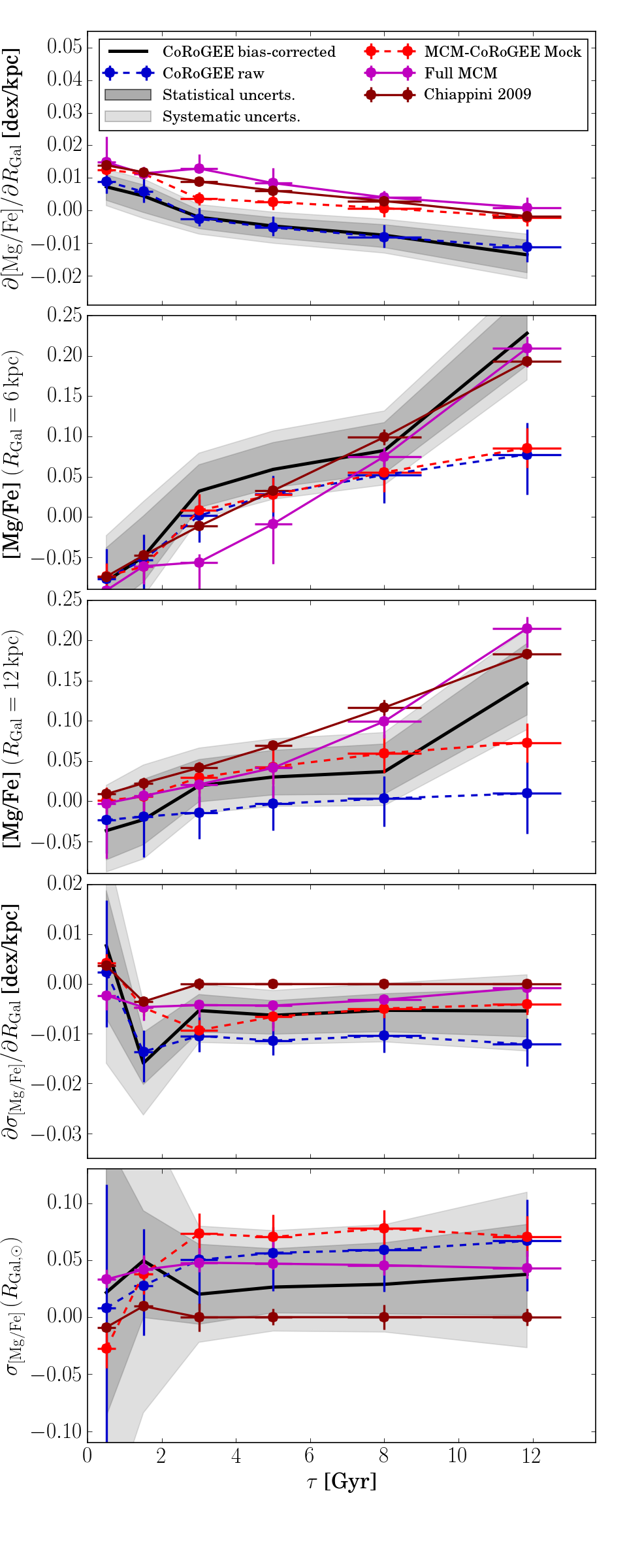}
 \caption{Results of our 4-parameter fits for the [Mg/Fe] vs. $R_{\rm Gal}$ distributions, in the same style as Fig. \ref{evolplot}.}
 \label{evolplot_mg}
\end{figure}

\section{The variation of radial [Mg/Fe] distributions with age}\label{mggrad}

Among many other elements, the APOGEE/ASPCAP pipeline also delivers Mg abundances for our red-giant sample. These can be compared to the predictions of chemical-evolution and chemo-dynamical models. In Fig. \ref{mgfegradient}, we show the [Mg/Fe] vs. $R_{\rm Gal}$ distributions for the same data and in the same age bins as in Fig. \ref{agegradient}. 

The behaviour of [Mg/Fe] with time and Galactocentric radius reflects the star-formation history of the Galactic disc. In an inside-out forming disc (such as the MCM model and its underlying GCE model; bottom row of Fig. \ref{mgfegradient}), the inner parts of the disc form more stars per unit of time than the outer parts, which results in a positive radial [Mg/Fe] gradient that slowly evolves with look-back time. Although the observational effects affecting our CoRoGEE sample smear out this clear signature in the [Mg/Fe]$-R_{\rm Gal}$ diagram (middle row of Fig. \ref{mgfegradient}), the data clearly confirm the inside-out formation of the thin disc.

While the [Mg/Fe] vs. $R_{\rm Gal}$ distributions from the pure chemical-evolution model seem to follow a quadratic rather than linear trend, the data in the range 6 kpc $<R_{\rm Gal}\lesssim$ 13 kpc can be well-fit with the same linear gradient + variable scatter model presented in Sec. \ref{grad}. We therefore also provide fits to this model in App. \ref{app}, Tab. \ref{gradienttable_mgfe_4}, and show the evolution of the fit parameters as a function of stellar age in Fig. \ref{evolplot_mg}.

As for the case of the radial [Fe/H] gradients, the MCM model matches the qualitative trends in the data extremely well, and also agrees with the data in an absolute sense in all age bins at a $2\sigma$ level. This is not a completely natural outcome, since the main observational constraints used to construct the GCE model were the present-day star-formation rate and [Fe/H] gradient, and the metallicity distribution in the solar neighbourhood. The comparison between the GCE and the MCM model (bottom panels of Fig. \ref{mgfegradient}) shows that the radial [Mg/Fe] distributions are not affected as severely by radial migration as the [Fe/H] gradients shown in Fig. \ref{agegradient}  (see also \citealt{Minchev2014b}). This is likely to be due to the shallow slope of the radial [Mg/Fe] gradient, and its slow evolution (flattening at greater ages). The [Mg/Fe] scatter in the MCM models decreases slightly with Galactocentric distance, an effect that is also seen in the data. The [Mg/Fe] scatter at the Solar radius in both model and (bias-corrected) data saturates already for ages $\gtrsim 2$ Gyr at a level of $\simeq 0.05$ dex.

An obvious drawback of our approach is that we do not explicitly fit the observationally-confirmed high-[$\alpha$/Fe] sequence (\citealt{Fuhrmann2011, Anders2014, Nidever2014}) separately, and therefore our thin-disc [Mg/Fe]-gradient measurement is slightly biased: High-[$\alpha$/Fe] stars with $|Z_{\mathrm{Gal}}|<0.3$ kpc, old or young, contribute to the scatter seen in the first row of Fig. \ref{mgfegradient}. While their number clearly increases with age, a small number of them is seen in the younger panels. As discussed in \citet{Chiappini2015a}, this is in disagreement with GCE predictions. It is still under debate whether these stars are truly young or either the product of close-binary evolution, or inaccurate mass measurements (\citealt{Martig2015, Brogaard2016, Jofre2016, Yong2016}). Our CoRoGEE results \citep{Chiappini2015a} appear to favour a physical origin, since the observed number of these peculiar objects is much higher in the inner-disc field LRc01 (at $R_{\rm Gal}<6$ kpc, for which $|Z_{\rm Gal}|>0.3$ kpc). 

\section{Conclusions}\label{concl}

We have used the wide coverage of the CoRoT-APOGEE sample in age and Galactocentric distance to study the age dependence of the Galactic radial [Fe/H] and [Mg/Fe] gradients in the range $\{6<R_{\rm Gal}\lesssim13$ kpc, $|Z_{\rm Gal}|<0.3$ kpc, 0.5 Gyr $\lesssim\tau\lesssim 13$ Gyr$\}$. When corrected for selection biases, we find that the slope of the [Fe/H] gradient is constant at a value of $\simeq-0.07$ dex/kpc in the age range $1<\tau< 4$ Gyr, and slightly flatter for the youngest age bin, in agreement with Cepheid results. For older ages (where our age measurements are more uncertain), the slope flattens to reach values compatible with a flat distribution. We further confirm that the mean metallicity in the Solar Neighbourhood has remained roughly constant at Solar values during the last $\sim 5$ Gyr. At the same time, the [Fe/H] abundance spread in the Solar Neighbourhood does not increase significantly with age, within our uncertainties, and stays around 0.15 dex for $\tau\gtrsim1$ Gyr. 

We have compared our results with mock observations of the chemo-dynamical Milky-Way model by \citet{Minchev2013, Minchev2014b}, and find a surprisingly good quantitative agreement for the [Fe/H]-$R_{\rm Gal}$, as well as for the [Mg/Fe]-$R_{\rm Gal}$ distributions. This enabled us to use the mock results to correct our measurements for selection biases; the bias-corrected results can be directly compared to other models as well as results from the literature. They agree well with previous estimates of the evolution of the radial metallicity gradient from the Geneva-Copenhagen survey \citep{Nordstroem2004, Casagrande2011}, while they disagree with the recent LAMOST study of \citet{Xiang2015} as well as most PN and OC results. These differences can be explained by systematic shifts in the distance and/or age scales for the case of the PNe and LAMOST turn-off stars, and by strong non-trivial selection biases and small-number statistics in the case of the OCs. 

We also investigate in more detail why, in the $5-8$ kpc region of the disc, intermediate-age OCs are found at larger metallicities than the youngest ones. Within the MCM model, we can explain this effect by strong radial mixing (due to both heating and migration) from the inner disc. Together with our proposition that non- and inward-migrating clusters are disrupted faster, the model can also explain the puzzling observation that the radial [Fe/H] gradient of the intermediate-age cluster population seems to be much steeper than both the young-population gradient and the gradient traced by intermediate-age field stars. 

Due to the high accuracy and precision of our distances and [Fe/H] estimates, in combination with sufficient statistics and easily-accountable selection biases, field-star studies like ours are likely to supersede gradient studies using open clusters or planetary nebulae as abundance tracers. Even though the absolute age scale of red-giant asteroseismology is not settled yet (e.g., \citealt{Brogaard2015, Brogaard2016, Miglio2016}), relative ages seem to be robustly determined for the vast majority of stars.
In the near future, many more fields in the K2 asteroseismic ecliptic-plane survey \citep{Howell2014} will be co-observed by the major spectroscopic stellar surveys (e.g., \citealt{Stello2015, Valentini2016a}). A joint spectroscopic and asteroseismic analysis of these fields will allow for a much larger sample, and consequently a more precise measurement of the Milky Way's chemo-dynamical history.

\begin{acknowledgements}
The authors thank the anonymous referee for her/his suggestions that helped improve the quality of the paper. FA thanks Omar Choudhury for the right tip at the right moment, and Martin Netopil for providing his open cluster data. TSR acknowledges support from CNPq-Brazil. BM acknowledges financial support from the ANR program IDEE Interaction Des \'Etoiles et des Exoplan\`etes. JM acknowledges support from the ERC Consolidator Grant funding scheme (project STARKEY, G.A. No. 615604). LG and TSR acknowledge partial support from PRIN INAF 2014 - CRA 1.05.01.94.05. TCB acknowledges partial support for this work from grant PHY 14-30152; Physics Frontier Center/JINA Center for the Evolution of the Elements (JINA-CEE), awarded by the US National Science Foundation. RAG received funding from the CNES PLATO grant at CEA. SM acknowledges support from the NASA grant NNX12AE17G. DAGH was funded by the Ram\'on y Cajal fellowship number RYC-2013-14182. DAGH and OZ acknowledge support provided by the Spanish Ministry of Economy and Competitiveness (MINECO) under grant AYA-2014-58082-P.

The CoRoT space mission, launched on December 27 2006, was developed and 
operated by CNES, with the contribution of Austria, Belgium, Brazil, ESA (RSSD 
and Science Program), Germany and Spain.\\

Funding for the SDSS-III Brazilian Participation Group has been provided by the 
Ministério de Ciência e Tecnologia (MCT), Funda\c{c}\~{a}o Carlos Chagas Filho 
de Amparo à Pesquisa do Estado do Rio de Janeiro (FAPERJ), Conselho Nacional de 
Desenvolvimento Científico e Tecnológico (CNPq), and Financiadora de Estudos e 
Projetos (FINEP). 
Funding for SDSS-III has been provided by the Alfred P. Sloan Foundation, the 
Participating Institutions, the National Science Foundation, and the U.S. 
Department of Energy Office of Science. The SDSS-III web site is 
\url{http://www.sdss3.org/}.\\
SDSS-III was managed by the Astrophysical Research Consortium for the 
Participating Institutions of the SDSS-III Collaboration including the 
University of Arizona, the Brazilian Participation Group, Brookhaven National 
Laboratory, Carnegie Mellon University, University of Florida, the French 
Participation Group, the German Participation Group, Harvard University, the 
Instituto de Astrofisica de Canarias, the Michigan State/Notre Dame/JINA 
Participation Group, Johns Hopkins University, Lawrence Berkeley National 
Laboratory, Max Planck Institute for Astrophysics, Max Planck Institute for 
Extraterrestrial Physics, New Mexico State University, New York University, 
Ohio State University, Pennsylvania State University, University of Portsmouth, 
Princeton University, the Spanish Participation Group, University of Tokyo, 
University of Utah, Vanderbilt University, University of Virginia, University 
of Washington, and Yale University.
\end{acknowledgements}

\bibliographystyle{aa}
\bibliography{FA_library}

\appendix
\section{Tabulated fit results}\label{app}

\begin{sidewaystable*}
\caption{Results of our 4-parameter fits (linear slope + variable scatter) to the [Fe/H] vs. $R_{\rm Gal}$ distributions of the various samples for 6 kpc $<R_{\rm Gal}<$ 15 kpc, in six age bins.}
\begin{tabular}{l cccccc}
Sample & $\tau < 1$ Gyr & $1 < \tau < 2$ Gyr  & $2 < \tau < 4$ Gyr  &  $4 < \tau < 6$ Gyr  & $6 < \tau < 10$ Gyr &  $ \tau > 10$ Gyr \\ 
\hline 
\hline 
 &&& $m$  [dex/kpc] &&& \\ 
\hline 
CoRoGEE raw fit & $-0.068^{+0.008}_{-0.008}$ & $-0.071^{+0.008}_{-0.008}$ & $-0.052^{+0.007}_{-0.007}$ & $-0.037^{+0.008}_{-0.009}$ & $-0.029^{+0.011}_{-0.011}$ & $-0.021^{+0.015}_{-0.016}$ \\ 
CoRoGEE bias-corrected & $-0.057^{+0.008}_{-0.008}\pm0.003$ & $-0.066^{+0.008}_{-0.008}\pm0.003$ & $-0.066^{+0.007}_{-0.007}\pm0.002$ & $-0.048^{+0.008}_{-0.009}\pm0.003$ & $-0.037^{+0.011}_{-0.011}\pm0.003$ & $-0.03^{+0.015}_{-0.016}\pm0.007$ \\ 
MCM-CoRoGEE mock & $-0.061^{+0.006}_{-0.006}$ & $-0.046^{+0.006}_{-0.006}$ & $-0.025^{+0.006}_{-0.006}$ & $-0.021^{+0.006}_{-0.006}$ & $-0.016^{+0.007}_{-0.007}$ & $-0.02^{+0.008}_{-0.008}$ \\ 
Full MCM model & $-0.053^{+0.002}_{-0.002}$ & $-0.049^{+0.003}_{-0.003}$ & $-0.048^{+0.003}_{-0.003}$ & $-0.041^{+0.003}_{-0.004}$ & $-0.033^{+0.005}_{-0.007}$& Not converged \\ 
Chiappini 2009 model & $-0.061^{+0.0}_{-0.0}$ & $-0.06^{+0.001}_{-0.001}$ & $-0.059^{+0.001}_{-0.001}$ & $-0.06^{+0.001}_{-0.001}$ & $-0.063^{+0.004}_{-0.004}$ & $-0.049^{+0.003}_{-0.003}$ \\ 
\hline 
 &&& $b$  [dex] &&& \\ 
\hline 
CoRoGEE raw fit & $0.62^{+0.08}_{-0.08}$ & $0.63^{+0.09}_{-0.09}$ & $0.43^{+0.07}_{-0.07}$ & $0.29^{+0.09}_{-0.08}$ & $0.18^{+0.11}_{-0.1}$ & $0.08^{+0.15}_{-0.14}$ \\ 
CoRoGEE bias-corrected & $0.54^{+0.08}_{-0.08}\pm0.03$ & $0.6^{+0.09}_{-0.09}\pm0.03$ & $0.65^{+0.07}_{-0.07}\pm0.02$ & $0.44^{+0.09}_{-0.08}\pm0.03$ & $0.19^{+0.11}_{-0.1}\pm0.03$ & $0.08^{+0.15}_{-0.14}\pm0.05$ \\ 
MCM-CoRoGEE mock & $0.59^{+0.06}_{-0.06}$ & $0.43^{+0.06}_{-0.06}$ & $0.14^{+0.06}_{-0.06}$ & $0.05^{+0.06}_{-0.06}$ & $-0.05^{+0.06}_{-0.07}$ & $-0.04^{+0.07}_{-0.07}$ \\ 
Full MCM model & $0.54^{+0.02}_{-0.02}$ & $0.48^{+0.02}_{-0.02}$ & $0.43^{+0.03}_{-0.03}$ & $0.27^{+0.04}_{-0.03}$ & $0.0^{+0.06}_{-0.04}$& Not converged \\ 
Chiappini 2009 model & $0.58^{+0.01}_{-0.01}$ & $0.52^{+0.01}_{-0.01}$ & $0.44^{+0.01}_{-0.01}$ & $0.34^{+0.01}_{-0.01}$ & $0.17^{+0.04}_{-0.04}$ & $-0.33^{+0.04}_{-0.04}$ \\ 
\hline 
 &&& $m_{\sigma}$  [dex/kpc] &&& \\ 
\hline 
CoRoGEE raw fit & $0.002^{+0.004}_{-0.004}$ & $-0.004^{+0.005}_{-0.005}$ & $-0.018^{+0.006}_{-0.005}$ & $-0.015^{+0.007}_{-0.006}$ & $-0.014^{+0.008}_{-0.007}$ & $-0.015^{+0.01}_{-0.01}$ \\ 
CoRoGEE bias-corrected & $0.016^{+0.004}_{-0.004}\pm0.004$ & $0.004^{+0.005}_{-0.005}\pm0.004$ & $-0.008^{+0.006}_{-0.005}\pm0.007$ & $0.001^{+0.007}_{-0.006}\pm0.005$ & $0.017^{+0.008}_{-0.007}\pm0.003$ & $0.016^{+0.01}_{-0.01}\pm0.006$ \\ 
MCM-CoRoGEE mock & $-0.004^{+0.004}_{-0.005}$ & $-0.005^{+0.004}_{-0.004}$ & $-0.013^{+0.004}_{-0.004}$ & $-0.014^{+0.004}_{-0.004}$ & $-0.011^{+0.005}_{-0.005}$ & $-0.005^{+0.007}_{-0.006}$ \\ 
Full MCM model & $-0.004^{+0.001}_{-0.001}$ & $-0.004^{+0.002}_{-0.002}$ & $-0.008^{+0.002}_{-0.002}$ & $-0.001^{+0.002}_{-0.002}$ & $0.008^{+0.004}_{-0.003}$& Not converged \\ 
Chiappini 2009 model & $0.0^{+0.001}_{-0.001}$ & $-0.0^{+0.001}_{-0.001}$ & $-0.0^{+0.001}_{-0.001}$ & $-0.0^{+0.001}_{-0.001}$ & $-0.0^{+0.004}_{-0.004}$ & $-0.0^{+0.006}_{-0.006}$ \\ 
\hline 
 &&& $b_{\sigma}$  [dex] &&& \\ 
\hline 
CoRoGEE raw fit & $0.12^{+0.04}_{-0.04}$ & $0.19^{+0.05}_{-0.05}$ & $0.36^{+0.05}_{-0.06}$ & $0.35^{+0.06}_{-0.06}$ & $0.35^{+0.07}_{-0.07}$ & $0.37^{+0.1}_{-0.09}$ \\ 
CoRoGEE bias-corrected & $-0.02^{+0.04}_{-0.04}\pm0.04$ & $0.13^{+0.05}_{-0.05}\pm0.04$ & $0.22^{+0.05}_{-0.06}\pm0.07$ & $0.13^{+0.06}_{-0.06}\pm0.06$ & $0.01^{+0.07}_{-0.07}\pm0.03$ & $0.02^{+0.1}_{-0.09}\pm0.07$ \\ 
MCM-CoRoGEE mock & $0.14^{+0.05}_{-0.04}$ & $0.16^{+0.04}_{-0.04}$ & $0.31^{+0.04}_{-0.04}$ & $0.32^{+0.04}_{-0.04}$ & $0.3^{+0.05}_{-0.05}$ & $0.22^{+0.06}_{-0.06}$ \\ 
Full MCM model & $0.13^{+0.01}_{-0.01}$ & $0.17^{+0.02}_{-0.02}$ & $0.23^{+0.02}_{-0.02}$ & $0.16^{+0.02}_{-0.02}$ & $0.11^{+0.03}_{-0.03}$& Not converged \\ 
Chiappini 2009 model & $0.0^{+0.01}_{-0.01}$ & $0.0^{+0.01}_{-0.01}$ & $0.0^{+0.01}_{-0.01}$ & $0.0^{+0.01}_{-0.02}$ & $0.0^{+0.04}_{-0.04}$ & $0.0^{+0.07}_{-0.07}$ \\ 
\hline 
\end{tabular}
\label{gradienttable_4}
\end{sidewaystable*}

\begin{sidewaystable*}
\caption{Results of our 4-parameter fits (linear slope + variable scatter) to the [O/H] vs. $R_{\rm Gal}$ distributions of the various samples for 6 kpc $<R_{\rm Gal}<$ 15 kpc, in six age bins.}
\begin{tabular}{l cccccc}
Sample & $\tau < 1$ Gyr & $1 < \tau < 2$ Gyr  & $2 < \tau < 4$ Gyr  &  $4 < \tau < 6$ Gyr  & $6 < \tau < 10$ Gyr &  $ \tau > 10$ Gyr \\ 
\hline 
\hline 
 &&& $m$  [dex/kpc] &&& \\ 
\hline 
CoRoGEE raw fit & $-0.051^{+0.006}_{-0.006}$ & $-0.055^{+0.006}_{-0.006}$ & $-0.045^{+0.005}_{-0.005}$ & $-0.035^{+0.006}_{-0.007}$ & $-0.03^{+0.008}_{-0.008}$ & $-0.024^{+0.01}_{-0.012}$ \\ 
CoRoGEE bias-corrected & $-0.047^{+0.006}_{-0.006}\pm0.001$ & $-0.055^{+0.006}_{-0.006}\pm0.002$ & $-0.055^{+0.005}_{-0.005}\pm0.002$ & $-0.039^{+0.006}_{-0.007}\pm0.002$ & $-0.034^{+0.008}_{-0.008}\pm0.003$ & $-0.028^{+0.01}_{-0.012}\pm0.004$ \\ 
MCM-CoRoGEE mock & $-0.039^{+0.004}_{-0.004}$ & $-0.035^{+0.003}_{-0.004}$ & $-0.023^{+0.005}_{-0.005}$ & $-0.017^{+0.005}_{-0.005}$ & $-0.017^{+0.005}_{-0.005}$ & $-0.022^{+0.006}_{-0.006}$ \\ 
Full MCM model & $-0.033^{+0.002}_{-0.002}$ & $-0.03^{+0.003}_{-0.003}$ & $-0.029^{+0.003}_{-0.004}$ & $-0.025^{+0.004}_{-0.005}$ & $-0.023^{+0.005}_{-0.006}$& Not converged \\ 
Chiappini 2009 model & $-0.033^{+0.001}_{-0.001}$ & $-0.034^{+0.001}_{-0.001}$ & $-0.041^{+0.002}_{-0.002}$ & $-0.047^{+0.001}_{-0.001}$ & $-0.056^{+0.002}_{-0.002}$ & $-0.052^{+0.003}_{-0.003}$ \\ 
\hline 
 &&& $b$  [dex] &&& \\ 
\hline 
CoRoGEE raw fit & $0.5^{+0.07}_{-0.07}$ & $0.52^{+0.06}_{-0.06}$ & $0.42^{+0.05}_{-0.05}$ & $0.35^{+0.06}_{-0.06}$ & $0.29^{+0.08}_{-0.07}$ & $0.22^{+0.12}_{-0.1}$ \\ 
CoRoGEE bias-corrected & $0.46^{+0.07}_{-0.07}\pm0.01$ & $0.52^{+0.06}_{-0.06}\pm0.02$ & $0.55^{+0.05}_{-0.05}\pm0.02$ & $0.42^{+0.06}_{-0.06}\pm0.01$ & $0.31^{+0.08}_{-0.07}\pm0.02$ & $0.24^{+0.12}_{-0.1}\pm0.04$ \\ 
MCM-CoRoGEE mock & $0.39^{+0.04}_{-0.04}$ & $0.33^{+0.04}_{-0.03}$ & $0.16^{+0.05}_{-0.04}$ & $0.08^{+0.05}_{-0.05}$ & $0.05^{+0.05}_{-0.05}$ & $0.07^{+0.06}_{-0.06}$ \\ 
Full MCM model & $0.31^{+0.02}_{-0.02}$ & $0.27^{+0.03}_{-0.02}$ & $0.25^{+0.04}_{-0.03}$ & $0.18^{+0.04}_{-0.03}$ & $0.08^{+0.05}_{-0.04}$& Not converged \\ 
Chiappini 2009 model & $0.32^{+0.01}_{-0.01}$ & $0.32^{+0.01}_{-0.01}$ & $0.35^{+0.02}_{-0.02}$ & $0.35^{+0.01}_{-0.01}$ & $0.31^{+0.02}_{-0.02}$ & $0.01^{+0.03}_{-0.03}$ \\ 
\hline 
 &&& $m_{\sigma}$  [dex/kpc] &&& \\ 
\hline 
CoRoGEE raw fit & $-0.002^{+0.003}_{-0.003}$ & $-0.006^{+0.004}_{-0.004}$ & $-0.013^{+0.004}_{-0.004}$ & $-0.009^{+0.005}_{-0.005}$ & $-0.01^{+0.005}_{-0.006}$ & $-0.011^{+0.008}_{-0.008}$ \\ 
CoRoGEE bias-corrected & $0.007^{+0.003}_{-0.003}\pm0.002$ & $-0.001^{+0.004}_{-0.004}\pm0.003$ & $-0.005^{+0.004}_{-0.004}\pm0.002$ & $0.003^{+0.005}_{-0.005}\pm0.002$ & $0.009^{+0.005}_{-0.006}\pm0.001$ & $0.008^{+0.008}_{-0.008}\pm0.006$ \\ 
MCM-CoRoGEE mock & $-0.011^{+0.004}_{-0.005}$ & $0.0^{+0.002}_{-0.002}$ & $-0.007^{+0.003}_{-0.003}$ & $-0.009^{+0.003}_{-0.003}$ & $-0.01^{+0.004}_{-0.004}$ & $-0.005^{+0.004}_{-0.005}$ \\ 
Full MCM model & $0.001^{+0.001}_{-0.001}$ & $0.005^{+0.001}_{-0.001}$ & $0.001^{+0.002}_{-0.002}$ & $0.006^{+0.003}_{-0.002}$ & $0.009^{+0.004}_{-0.003}$& Not converged \\ 
Chiappini 2009 model & $0.006^{+0.001}_{-0.001}$ & $-0.007^{+0.001}_{-0.001}$ & $-0.006^{+0.001}_{-0.001}$ & $0.004^{+0.001}_{-0.001}$ & $0.0^{+0.002}_{-0.003}$ & $0.0^{+0.007}_{-0.007}$ \\ 
\hline 
 &&& $b_{\sigma}$  [dex] &&& \\ 
\hline 
CoRoGEE raw fit & $0.14^{+0.04}_{-0.03}$ & $0.18^{+0.04}_{-0.04}$ & $0.27^{+0.04}_{-0.04}$ & $0.25^{+0.04}_{-0.04}$ & $0.27^{+0.05}_{-0.05}$ & $0.28^{+0.07}_{-0.07}$ \\ 
CoRoGEE bias-corrected & $0.05^{+0.04}_{-0.03}\pm0.02$ & $0.14^{+0.04}_{-0.04}\pm0.02$ & $0.16^{+0.04}_{-0.04}\pm0.03$ & $0.11^{+0.04}_{-0.04}\pm0.02$ & $0.06^{+0.05}_{-0.05}\pm0.01$ & $0.07^{+0.07}_{-0.07}\pm0.07$ \\ 
MCM-CoRoGEE mock & $0.15^{+0.05}_{-0.04}$ & $0.07^{+0.02}_{-0.02}$ & $0.2^{+0.03}_{-0.03}$ & $0.23^{+0.03}_{-0.03}$ & $0.24^{+0.04}_{-0.03}$ & $0.18^{+0.04}_{-0.04}$ \\ 
Full MCM model & $0.03^{+0.01}_{-0.01}$ & $0.02^{+0.01}_{-0.01}$ & $0.07^{+0.02}_{-0.02}$ & $0.04^{+0.02}_{-0.02}$ & $0.03^{+0.03}_{-0.03}$& Not converged \\ 
Chiappini 2009 model & $-0.05^{+0.01}_{-0.01}$ & $0.06^{+0.01}_{-0.01}$ & $0.06^{+0.01}_{-0.01}$ & $-0.04^{+0.01}_{-0.01}$ & $-0.0^{+0.03}_{-0.03}$ & $-0.0^{+0.08}_{-0.08}$ \\ 
\hline 
\end{tabular}
\label{gradienttable_oh_4}
\end{sidewaystable*}

\begin{sidewaystable*}
\caption{Results of our 4-parameter fits (linear slope + variable scatter) to the [Mg/Fe] vs. $R_{\rm Gal}$ distributions of the various samples for 6 kpc $<R_{\rm Gal}<$ 15 kpc, in six age bins.}
\begin{tabular}{l cccccc}
Sample & $\tau < 1$ Gyr & $1 < \tau < 2$ Gyr  & $2 < \tau < 4$ Gyr  &  $4 < \tau < 6$ Gyr  & $6 < \tau < 10$ Gyr &  $ \tau > 10$ Gyr \\ 
\hline 
\hline 
 &&& $m$  [dex/kpc] &&& \\ 
\hline 
CoRoGEE raw fit & $0.009^{+0.004}_{-0.004}$ & $0.006^{+0.003}_{-0.005}$ & $-0.003^{+0.002}_{-0.003}$ & $-0.005^{+0.003}_{-0.004}$ & $-0.008^{+0.003}_{-0.004}$ & $-0.011^{+0.005}_{-0.005}$ \\ 
CoRoGEE bias-corrected & $0.007^{+0.004}_{-0.004}\pm0.002$ & $0.004^{+0.003}_{-0.005}\pm0.002$ & $0.004^{+0.002}_{-0.003}\pm0.003$ & $-0.003^{+0.003}_{-0.004}\pm0.001$ & $-0.008^{+0.003}_{-0.004}\pm0.002$ & $-0.014^{+0.005}_{-0.005}\pm0.002$ \\ 
MCM-CoRoGEE mock & $0.012^{+0.002}_{-0.002}$ & $0.011^{+0.002}_{-0.002}$ & $0.004^{+0.002}_{-0.002}$ & $0.003^{+0.002}_{-0.002}$ & $0.001^{+0.003}_{-0.003}$ & $-0.002^{+0.003}_{-0.003}$ \\ 
Full MCM model & $0.015^{+0.001}_{-0.008}$ & $0.011^{+0.001}_{-0.002}$ & $0.013^{+0.001}_{-0.004}$ & $0.008^{+0.001}_{-0.005}$ & $0.004^{+0.001}_{-0.002}$ & $0.001^{+0.002}_{-0.003}$ \\ 
Chiappini 2009 model & $0.014^{+0.001}_{-0.001}$ & $0.012^{+0.001}_{-0.001}$ & $0.009^{+0.0}_{-0.0}$ & $0.006^{+0.0}_{-0.0}$ & $0.003^{+0.001}_{-0.001}$ & $-0.002^{+0.001}_{-0.001}$ \\ 
\hline 
 &&& $b$  [dex] &&& \\ 
\hline 
CoRoGEE raw fit & $-0.13^{+0.04}_{-0.04}$ & $-0.09^{+0.05}_{-0.03}$ & $0.02^{+0.03}_{-0.02}$ & $0.06^{+0.04}_{-0.03}$ & $0.1^{+0.04}_{-0.03}$ & $0.15^{+0.05}_{-0.04}$ \\ 
CoRoGEE bias-corrected & $-0.12^{+0.04}_{-0.04}\pm0.02$ & $-0.08^{+0.05}_{-0.03}\pm0.02$ & $-0.08^{+0.03}_{-0.02}\pm0.03$ & $0.02^{+0.04}_{-0.03}\pm0.02$ & $0.13^{+0.04}_{-0.03}\pm0.02$ & $0.31^{+0.05}_{-0.04}\pm0.02$ \\ 
MCM-CoRoGEE mock & $-0.15^{+0.02}_{-0.02}$ & $-0.13^{+0.02}_{-0.02}$ & $-0.01^{+0.02}_{-0.02}$ & $0.01^{+0.02}_{-0.02}$ & $0.05^{+0.03}_{-0.03}$ & $0.1^{+0.03}_{-0.03}$ \\ 
Full MCM model & $-0.18^{+0.07}_{-0.01}$ & $-0.13^{+0.02}_{-0.01}$ & $-0.13^{+0.04}_{-0.01}$ & $-0.06^{+0.05}_{-0.01}$ & $0.05^{+0.02}_{-0.01}$ & $0.2^{+0.03}_{-0.02}$ \\ 
Chiappini 2009 model & $-0.16^{+0.01}_{-0.01}$ & $-0.12^{+0.01}_{-0.01}$ & $-0.06^{+0.0}_{-0.01}$ & $-0.0^{+0.0}_{-0.0}$ & $0.08^{+0.01}_{-0.01}$ & $0.2^{+0.01}_{-0.01}$ \\ 
\hline 
 &&& $m_{\sigma}$  [dex/kpc] &&& \\ 
\hline 
CoRoGEE raw fit & $0.002^{+0.011}_{-0.015}$ & $-0.014^{+0.006}_{-0.004}$ & $-0.01^{+0.003}_{-0.004}$ & $-0.011^{+0.003}_{-0.004}$ & $-0.01^{+0.003}_{-0.004}$ & $-0.012^{+0.004}_{-0.005}$ \\ 
CoRoGEE bias-corrected & $0.008^{+0.011}_{-0.015}\pm0.009$ & $-0.016^{+0.006}_{-0.004}\pm0.006$ & $-0.019^{+0.003}_{-0.004}\pm0.023$ & $-0.014^{+0.003}_{-0.004}\pm0.017$ & $-0.005^{+0.003}_{-0.004}\pm0.002$ & $-0.005^{+0.004}_{-0.005}\pm0.003$ \\ 
MCM-CoRoGEE mock & $0.004^{+0.002}_{-0.002}$ & $-0.005^{+0.002}_{-0.002}$ & $-0.009^{+0.002}_{-0.002}$ & $-0.007^{+0.002}_{-0.002}$ & $-0.005^{+0.002}_{-0.002}$ & $-0.004^{+0.002}_{-0.002}$ \\ 
Full MCM model & $-0.002^{+0.003}_{-0.001}$ & $-0.005^{+0.003}_{-0.001}$ & $-0.004^{+0.002}_{-0.001}$ & $-0.004^{+0.005}_{-0.001}$ & $-0.003^{+0.001}_{-0.001}$ & $-0.001^{+0.003}_{-0.001}$ \\ 
Chiappini 2009 model & $0.004^{+0.0}_{-0.0}$ & $-0.004^{+0.0}_{-0.0}$ & $0.0^{+0.001}_{-0.001}$ & $-0.0^{+0.001}_{-0.001}$ & $-0.0^{+0.001}_{-0.001}$ & $-0.0^{+0.001}_{-0.001}$ \\ 
\hline 
 &&& $b_{\sigma}$  [dex] &&& \\ 
\hline 
CoRoGEE raw fit & $-0.01^{+0.14}_{-0.12}$ & $0.14^{+0.05}_{-0.05}$ & $0.14^{+0.05}_{-0.03}$ & $0.15^{+0.04}_{-0.03}$ & $0.15^{+0.04}_{-0.03}$ & $0.17^{+0.05}_{-0.04}$ \\ 
CoRoGEE bias-corrected & $-0.04^{+0.14}_{-0.12}\pm0.1$ & $0.18^{+0.05}_{-0.05}\pm0.09$ & $0.21^{+0.05}_{-0.03}\pm0.23$ & $0.16^{+0.04}_{-0.03}\pm0.17$ & $0.07^{+0.04}_{-0.03}\pm0.02$ & $0.08^{+0.05}_{-0.04}\pm0.03$ \\ 
MCM-CoRoGEE mock & $-0.06^{+0.02}_{-0.02}$ & $0.08^{+0.02}_{-0.02}$ & $0.15^{+0.02}_{-0.02}$ & $0.13^{+0.02}_{-0.02}$ & $0.12^{+0.02}_{-0.02}$ & $0.1^{+0.02}_{-0.02}$ \\ 
Full MCM model & $0.05^{+0.01}_{-0.01}$ & $0.08^{+0.01}_{-0.01}$ & $0.08^{+0.01}_{-0.01}$ & $0.08^{+0.01}_{-0.02}$ & $0.07^{+0.01}_{-0.01}$ & $0.05^{+0.01}_{-0.02}$ \\ 
Chiappini 2009 model & $-0.04^{+0.0}_{-0.0}$ & $0.04^{+0.0}_{-0.0}$ & $-0.0^{+0.01}_{-0.01}$ & $0.0^{+0.01}_{-0.01}$ & $0.0^{+0.01}_{-0.01}$ & $0.0^{+0.01}_{-0.01}$ \\ 
\hline 
\end{tabular}
\label{gradienttable_mgfe_4}
\end{sidewaystable*}

\end{document}